\pdfoutput=1  % To generate pdf without going through dvi->ps->pdf (pdflatex)
\documentclass[12pt]{iopart}

\usepackage{iopams}
\usepackage[numbers]{natbib}
\usepackage{graphicx,bm,calc}

\usepackage[usenames,dvipsnames]{color}

\usepackage{hyperref}
\hypersetup{%
  breaklinks = {true},
  citecolor = {blue},
  colorlinks = {true},
  linkcolor = {red},
  pdfauthor = {\textcopyright\ Gareth W. Jones},
  pdfcreator = {\LaTeX\ and \flqq hyperref\frqq},
}

\graphicspath{ {../Figs/} } % Search path for includegraphics+graphicx
\newcommand{\pd}[2]{\frac{\partial#1}{\partial#2}}
\newcommand{\mb}[1]{\boldsymbol{#1}}
\newcommand{\mc}[1]{\mathcal{#1}}
\renewcommand{\mr}[1]{\mathrm{#1}}
\newcommand{\bra}[1]{\left(#1\right)}
\newcommand{\sbra}[1]{\left[#1\right]}
\newcommand{\at}[2]{\left. #1 \right|_{#2}}

\newcommand{\Om}{\Omega}
\newcommand{\al}{\alpha}
\newcommand{\be}{\beta}
\newcommand{\ga}{\gamma}
\newcommand{\de}{\delta}
\newcommand{\ep}{\varepsilon}

\newcommand{\ka}{\kappa}
\newcommand{\la}{\lambda}
\newcommand{\rh}{\rho}
\newcommand{\im}{\mathrm{i}}
\newcommand{\Bb}{\bm{B}}
\renewcommand{\br}{\bm{r}}

\newcommand{\bX}{\bm{X}}
\newcommand{\Acal}{\mathcal{A}}
\newcommand{\Eqref}[1]{equation~(\ref{#1})}
\newcommand{\appref}[1]{\ref{#1}}
\begin{document}

\title[Designing electronic properties of 2D crystals\ldots]{%
Designing electronic properties of two-dimensional crystals through 
optimization of deformations%
}
\author{%
Gareth W. Jones$^{1,2}$%
  \footnote{Current affiliation: The University of Manchester} 
and Vitor M. Pereira$^2$
}
\address{%
$^1$School of Mathematics, The University of Manchester, 
Manchester M13 9PL, England \\
$^2$Graphene Research Centre \& Department of Physics.  
National University of Singapore, 2 Science Drive 3, Singapore 117542
}
\eads{%
$^1$gareth.jones-10@manchester.ac.uk, 
$^2$vpereira@nus.edu.sg
}

\begin{abstract}
One of the enticing features common to most of the two-dimensional electronic 
systems that, in the wake of (and in parallel with) graphene, are currently at 
the forefront of materials science research is the ability to easily introduce 
a combination of planar deformations and bending in the system. 
Since the electronic properties are ultimately determined by the details of 
atomic orbital overlap, such mechanical manipulations translate into modified 
(or, at least, perturbed) electronic properties.
Here, we present a general-purpose optimization framework for tailoring 
physical properties of two-dimensional electronic systems by manipulating the 
state of local strain, allowing a one-step route from their design to 
experimental implementation.
A definite example, chosen for its relevance in light of current experiments in 
graphene nanostructures, is the optimization of the experimental parameters 
that generate a prescribed spatial profile of pseudomagnetic fields in graphene.
But the method is general enough to accommodate a multitude of possible 
experimental parameters and conditions whereby deformations can be imparted to 
the graphene lattice, and complies, by design, with graphene's elastic 
equilibrium and elastic compatibility constraints.
As a result, it efficiently answers the inverse problem of determining the 
optimal values of a set of external or control parameters (such as substrate 
topography, sample shape, load distribution, etc.) that result in a graphene 
deformation whose associated pseudomagnetic field profile best matches a 
prescribed target.
The ability to address this inverse problem in an expedited way is one key step 
for practical implementations of the concept of two-dimensional systems with 
electronic properties strain-engineered to order.
The general-purpose nature of this calculation strategy means that it can be 
easily applied to the optimization of other relevant physical quantities 
which directly depend on the local strain field, not just in graphene but in 
other two-dimensional electronic membranes.
\end{abstract}


% 81.05.ue Carbon-based materials - graphene, 
% 02.30.Zz Inverse problems
% 46.25.-y Elasticity in continuum mechanics of solids
% 46.70.De Mechanical properties beams, plates, and shells
% 73.22.Pr: graphene: electronic structure

\maketitle

% \begin{bibunit}[unsrt]

\section{Introduction}

With their intrinsic two-dimensionality, ``electronic membranes'' are easily 
pulled or pinched by atomic-scale tips 
\cite{LeeSCIENCE2008,Klimov-Jung-etal-2012,Xu-Yang-etal-2012}, can be made to 
conform to the substrate topography 
\cite{Metzger2010Biaxial,Scharfenberg2011Probing,Tomori-Kanda-etal-2011}, can 
be inflated as balloons \cite{Koenig-Boddeti-etal-2011}, can be 
stretched \cite{Bao-Miao-etal-2009} or bent \cite{Georgiou-Britnell-etal-2011}, 
crumpled on demand \cite{Zang2013Multifunctionality}, and so on. 
Hence, two-dimensional crystals are an excellent case (and opportunity) of 
correlation between electronic behavior and shape with tremendous implications 
in bridging \emph{soft} and \emph{hard} condensed matter.
For example, if a physical property is sensitive to the state of deformation of 
the system it can be used to monitor its shape, strain, etc.; conversely, the 
shape variables can be manipulated so that the physical quantity in question 
behaves in a desired way, has a certain magnitude, or a particularly useful 
spatial profile. In addition, the fact that some of these two-dimensional 
electronic membranes can be easily, and non-detrimentally, embedded in living 
tissues, organs or plants 
\cite{Schmidt2012Bioelectronics,Mannoor2012Graphenebased}, brings the 
tantalizing prospect of using them in bioelectronics. The method to be 
discussed next can be a valuable tool there, in the cases where the system's 
functionality is determined by the shape or deformation state of the membrane.

To be specific---but by no means implying a limitation in scope---consider 
the problem of strained graphene. 
It is well-established that a mechanically-strained graphene sheet is very 
resilient \cite{LeeSCIENCE2008} even in polycrystaline form 
\cite{Rasool2013Measurement,Zhang2014Fracture}, and has altered 
electronic transport properties. In particular, and among other features, it
exhibits an unconventional contribution in the electron--phonon coupling leading 
to the emergence of so-called pseudomagnetic fields (PMF) 
\cite{Kane:1997,Suzuura:2002,CastroNeto-Guinea-etal-2009, 
Vozmediano-Katsnelson-Guinea-2010}. These fields appear naturally in the 
effective (low-energy) description of the electronic problem in deformed 
graphene, and are a consequence of the peculiar lattice structure. 
Briefly, the celebrated Weyl--Dirac equation that captures most of the 
electronic phenomenology of graphene ($\mathcal{H} = 
v_F\,\bm{p}\cdot\bm{\sigma}$ for one of the $K$ points in the Brillouin zone) 
is corrected in the presence of lattice deformations in a way that amounts to 
substituting $\bm{p}\to\bm{p}+e\bm{\mathcal{A}}$, where $\bm{\mathcal{A}}$ 
encodes all the details of the deformation and how it perturbs the electronic 
hopping amplitudes (defined below) \cite{Kane:1997}. 
As a result, even though $\bm{\mathcal{A}}$ is not a magnetic vector potential, 
the actual dynamics has the same characteristics and the Dirac electrons in 
graphene react to static and non-uniform lattice deformations as though they 
were under the influence of an effective magnetic field, with all the 
consequences that a magnetic field brings to electronic motion, except that 
time-reversal symmetry is not broken and, thus, $\bm{\mathcal{A}}$ will have an 
opposite sign for the effective Hamiltonian at the time-reversal transformed 
$K'$ point. One such consequence is the modification of the electronic energy 
spectrum with the development of local Landau levels for certain lattice 
deformations. 
This has been recently confirmed by local scanning tunneling spectroscopy (STS) 
measurements on nanometer-scale graphene nano-blisters which revealed Landau 
level resonances associated with PMFs in the range $300$--$600$\,T 
\cite{Levy-Burke-etal-2010,LuNatureComm2012}. Equally interesting
space-dependent Fermi velocities have also been reported in recent experiments 
on strained graphene \cite{Jang2014Observation}, bringing this other theoretical 
prediction \cite{deJuan2012Space,deJuan2013Gauge,BarrazaLopez-Discrete} and 
implication of non-uniform deformation fields  closer to reality.

The possibilities associated with these discoveries and the confirmation of the 
drastic impact that moderate lattice deformations can have in graphene's 
electronic spectrum have spurred researchers to investigate deformation modes 
allowing a degree of control over PMFs that can be tailored for specific ends, 
such as electronic confinement, guiding, and so on. This is a concept known as 
strain engineering or straintronics 
\citep{Pereira-CastroNeto-2009,Guinea-Katsnelson-Geim-2009,
Tomori-Kanda-etal-2011,Lim-Koon-etal-2012,Guinea-2012}.

\begin{figure*}
  \centering
  \includegraphics[width=0.7\textwidth]{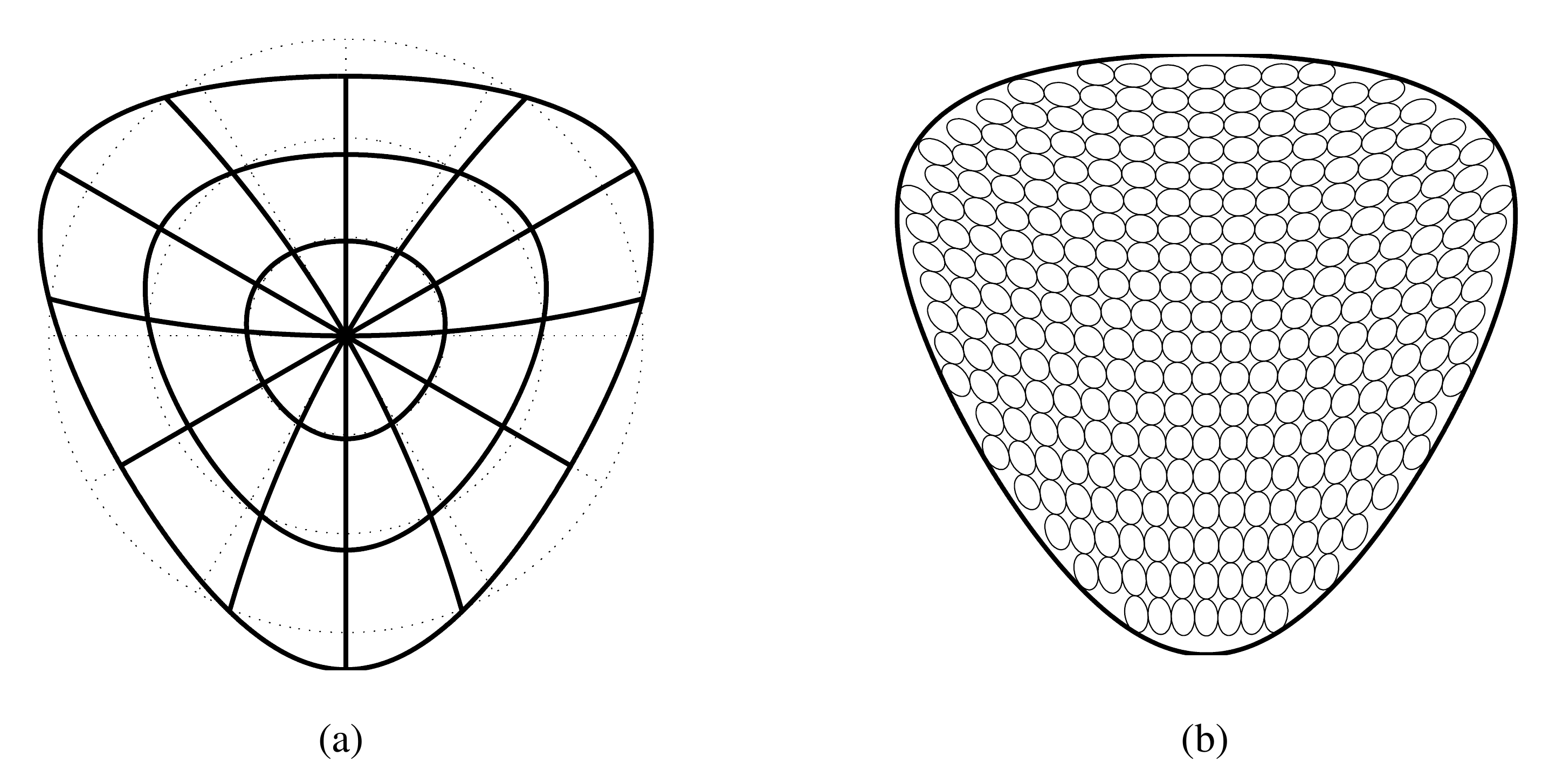}%
  \caption{
    The two dimensional solution of \citet{Guinea-Katsnelson-Geim-2009}. 
    (a) An initially circular and isotropic graphene sheet is 
    deformed to a rounded triangular shape. 
    (b) The magnitude and direction of local stretch is 
    indicated by the ellipses, which are the images under the deformation of 
    small circles in the undeformed sheet.
  }
  \label{fig:guinea}
\end{figure*}

Since the electronic dynamics can be straightforwardly determined once a 
space-dependent (pseudo)\-magnetic field $B(\bX)$ is prescribed, and since much 
is already known about the behavior of Dirac electrons in graphene under the 
influence of magnetic field profiles such as barriers, wells, channels, and so 
on, it is natural to approach this strain-engineering problem from the 
perspective of seeking which deformation fields applied to the carbon 
lattice lead to that prescribed PMF profile. As will be clear in subsequent 
sections, the solution is not unique. If not for anything else, this should be 
clear from the fact that there is a ``gauge'' freedom in selecting the vector 
potential $\bm{\Acal}$ from $\Bb=\nabla\times\bm{\Acal}$. The simplest of such 
problems is to determine which displacement fields lead to a strictly uniform 
(space-independent) $B$. The first notable theoretical investigation along 
these lines was that of \citet{Guinea-Katsnelson-Geim-2009}, who restricted 
their analysis to deformations in the plane. In this regime the PMFs are linear 
in the displacement field, allowing one to calculate an in-plane deformation 
field giving rise to any given PMF. In particular, to generate a constant or 
mostly constant PMF requires a characteristic deformation with 3-fold symmetry 
(see \Fref{fig:guinea}), and the magnitude of the resulting PMF depends 
explicitly on the relative orientation of the deformation field and the 
underlying graphene lattice. This particular strain configuration has been 
recently explored in experiments on ``artificial graphene'' 
\cite{Gomes2012Designer}.

Extending deformations to three dimensions introduces nonlinearity into the 
strain field, and such simple solutions are no longer available.
Continuum mechanical theoretical investigations in simple geometries such as 
one-dimensional bending \citep{Guinea-Geim-etal-2010,low_strain-induced_2010} 
or radially-symmetric bubbles \citep{Kim-Blanter-Ahn-2011}, and atomistic 
simulations of graphene sheets adhered to nanoscale patterned substrates 
\citep{NeekAmal-Covaci-Peeters-2012,NeekAmal-Peeters-2012,Barraza-Lopez:2013} 
are examples of 
\emph{forward problems}: calculating the PMF associated with a certain 
deformation. But such approaches are unlikely to solve the \emph{inverse 
problem} of finding the deformation mode required to produce a given PMF. In 
addition, given the current surge of experimental interest in 
deliberately inducing non-uniform strains in graphene, various possible routes 
are being explored 
\cite{Tomori-Kanda-etal-2011,LuNatureComm2012,Li2012Nanoscale,
Shioya2014Straining}. 
To be experimentally relevant, an attempt to effectively tackle the crucial 
inverse problem should be generic enough to encompass such diverse means to 
experimentally generate the desired strain fields.

\begin{figure*}
  \centering
  \includegraphics[width=0.7\textwidth]{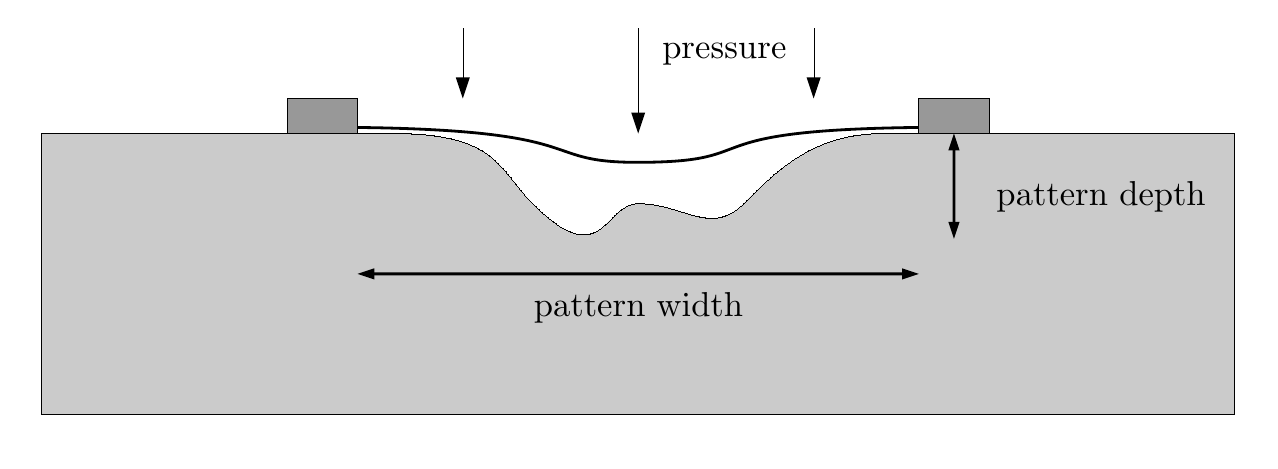}%
  \caption{
    Experimental situation to be modeled and taken throughout this report as a 
    practical example of our method to solve the inverse strain-engineering 
    problem in graphene.
    Graphene is placed on a patterned substrate with which it interacts via 
    Van~der~Waals forces. Hydrostatic pressure and substrate profile are the 
    two 
    control parameters here, and the former is used to control the degree of
    conformation of graphene the substrate.
  }
  \label{fig:expt}
\end{figure*}

This report presents a general-purpose framework which may be used to solve 
such inverse problems in graphene. In particular, for a given target PMF and 
experimental configuration, the method aims to find the optimal 
\emph{deformation control} that, when applied to the graphene sheet, produces 
the desired PMF. \emph{Desired PMF} refers to any specified space dependence 
of $B(\bX)$. 
\emph{Deformation control} is the name for the geometric and 
mechanical parameters of the experiment that may be varied to change the 
deformation field. In the 2d example of \citet{Guinea-Katsnelson-Geim-2009} 
the deformation control is the displacement field applied to the outer 
boundary. 
In the case of graphene adhered to a patterned substrate the shape of the 
substrate performs that role. In this particular setup, which we will use 
extensively as an illustrative example in this report, the graphene sheet is 
assumed to have been transferred onto a patterned substrate, and forced to 
conform to its shape by combined hydrostatic pressure and adhesion forces (see 
\Fref{fig:expt}). The aim in this case is to find the substrate pattern and 
pressure (the two deformation controls for this example) for which the deformed 
graphene sheet exhibits a desired target PMF. But we underline that the 
approach is straightforwardly applicable to any other target quantity with a 
known dependence on the strain field.

We begin by summarizing the elastic properties of graphene and the elastic 
plate equations that govern its deformation when considered as a continuum 
elastic membrane. We then discuss the coupling of local deformations to the 
electronic degrees of freedom by means of the PMF and the optimization 
framework that forms the basis of our solution method. This is followed by a 
summary of the numerical algorithm used to solve the problem and, 
finally, as an example calculation, we present the computed substrate shapes 
that generate various predefined PMFs in an overlaid graphene sheet, and 
discuss the versatility of our framework for application in different 
experimental and theoretical scenarios well beyond the example calculations 
shown here for illustration. For completeness, various technical considerations 
and details are included as appendices to the main text.

\section{Graphene as an elastic electronic continuum}

\subsection{Graphene's elastic parametrization}\label{sect:elastic-basics}

The deformation of graphene is modeled using the equations from continuum 
elasticity. This formulation is chosen for its applicability across a wide range 
of lengthscales. In spite of recent developments based on discrete 
differential geometry to directly relate atomistic configurations with 
electronic properties of the type we envision \cite{BarrazaLopez-Discrete}, 
an atomistic approach to the elastic relaxation problem becomes easily 
unfeasible at scales of a few nanometers due to the intrinsically more 
numerically demanding nature of inverse problems.

The deformation of a graphene sheet is thus described in terms of its deviation 
from a flat two-dimensional surface. The point $\bX=(X,Y,0)$ is 
transformed to $\br=(x_1,x_2,x_3)=(x,y,z)$ in three-dimensional space, where 
$x_\al = X_\al + v_\al(X_1,X_2)$, and $z=w(X,Y)$.
The deformation measures which describe the sheet's local stretching and 
bending are respectively the strain and curvature tensors $\ep_{\al\be}$ and 
$\rho_{\al\be}$.
Since these are complicated to write in terms of the displacement components, 
in practice simplified forms are used (for completeness, a detailed discussion 
of the form of $\ep_{\al\be}$ and $\rho_{\al\be}$ is included in 
\appref{ap:elasticity}).
The most common simplification is perhaps the von K\'arm\'an 
approximation, which uses the expressions
\begin{eqnarray}
  \ep_{\al\be}=\frac{1}{2}\bra{\frac{\partial v_\al}{\partial X_\be}+\frac{\partial v_\be}{\partial X_\al}+\frac{\partial w}{\partial X_\al}\frac{\partial w}{\partial X_\be}}
  ,\qquad
  \rho_{\al\be}=\frac{\partial^2 w}{\partial X_\al\partial X_\be} .
  \label{curvature-vk}
\end{eqnarray}
The stress and moment resultants are assumed to be isotropic and linear in the 
strain and curvature tensors: %
\begin{eqnarray}
N_{\al\be}=C\,\mathbb{A}_{\al\be\ga\de} \; \ep_{\ga\de},\qquad 
M_{\al\be}=D\,\mathbb{A}_{\al\be\ga\de} \; \rh_{\ga\de},\label{NM-def}
\end{eqnarray}
where we have used Einstein's summation convention, and
\begin{eqnarray}
\mathbb{A}_{\al\be\ga\de}=\bra{\frac{1-\nu}{2}}(\de_{\al\ga}\de_{\be\de}+\de_{
\al\de}\de_{\be\ga}) +\nu\de_{\al\be}\de_{\ga\de},
\end{eqnarray}
$\nu$ being the Poisson ratio and $C$, $D$ the stretching and bending moduli, 
respectively.
To calculate the stretching modulus $C$, we use the results of 
\citet{Wei-Fragneaud-etal-2009} which, in our notation, are 
\begin{eqnarray}
  C=358.1\,\mathrm{N\,m^{-1}},\qquad 
  C\nu=60.4\,\mathrm{N\,m^{-1}}\qquad\Rightarrow\qquad\nu=0.169.
\end{eqnarray}
This value of $\nu$ agrees with the experimentally-measured Poisson ratio in 
graphite \cite{Blakslee:1970}. The value used for the bending modulus was 
that of \citet{Kudin-Scuseria-Yakobson-2001}, calculated \emph{ab initio} as 
$D=1.46\,\textrm{eV}=2.34\times10^{-19}\,\textrm{N\,m}$. 

Note that the moduli $C$ and $D$ are independent in our formulation of the 
elastic response of graphene, which is treated as a purely two-dimensional 
sheet. This means that thickness \emph{is not} a parameter in our modeling. 
We emphasize this aspect because graphene's elasticity is often modeled by 
treating it as a three-dimensional material which is thin in one dimension, 
\emph{i.e.} a conventional elastic thin plate. In those cases the stiffness and 
bending moduli are often written in terms of the three-dimensional Young's 
modulus $E$ and the thickness $h$:
\begin{eqnarray}
  C=\frac{Eh}{1-\nu^2},\qquad D=\frac{Eh^3}{12(1-\nu^2)},
\end{eqnarray}
where a typical value $h\approx0.3\,\textrm{nm}$ for graphene's effective 
thickness \cite{Huang:2006} is used. For example such expressions have been 
used to cite graphene's Young's modulus as of the order of $1\,\textrm{TPa}$ 
\cite{Lu:1997,Hernandez:1998,LeeSCIENCE2008}. 
While this may be useful to convey the scale and exceptional strength of 
graphene, the same numbers lead to an inaccurate value for the bending modulus 
$D$. Treating graphene as a continuous 3D elastic object might be a convenient 
approximation, but in keeping with graphene's two-dimensional nature, we 
retain the parameters $C$ and $D$ as our main quantities here rather than 
express them in terms of Young's modulus $E$.

\subsection{Equilibrium conditions}\label{sect:elastic-eqm}

In addition to the kinematic and constitutive equations for a sheet of 
graphene, one must establish the equations of force balance to close the system.
These are typically found by minimizing the potential energy functional 
consisting of two terms:\ $\mc{E}^\mathrm{elast}$, the stored elastic energy, 
and $\mc{E}^\mathrm{ext}$, the potential energy associated with external forces 
applied to the sheet.
The latter may be surface tractions or adhesive forces (for simplicity we 
neglect any forces explicitly applied to the edge of the graphene sheet).
The two energy terms are given by
\numparts
\begin{eqnarray}
  \mc{E}^\mathrm{elast}[w,v_1,v_2] & = 
  \int_\Omega\bra{\frac{1}{2}\ep_{\al\be}N_{\al\be} + 
  \frac{1}{2}\rh_{\al\be}M_{ \al\be}}\mr{d}^2\mb{X},\label{eq:energy1}\\
  \mc{E}^\mathrm{ext}[w,v_1,v_2;\la_i] & = 
    \int_\Omega V[w,v_1,v_2;\la_i]\,\mr{d}^2\mb{X}
  .
  \label{eq:energy2}
\end{eqnarray}
\endnumparts
For now the energy density of external forces, $V[w,v_1,v_2;\la_i]$, is left 
unspecified.
However, we do note that it is in this term that the influence of the control 
variables $\la_i$ is encoded; these may include, for instance, a 
parametrization of an underlying substrate, or the components of a surface 
traction field. The specific example corresponding to \Fref{fig:expt} will be 
presented in detail later.

In the standard variational formulation of the problem the potential energy is 
minimized by setting its first variation to zero, giving us three weak form 
equations for $v_1$, $v_2$, and $w$.
However, this requires some regularity in the behavior of the transverse 
displacement $w$:\ its first derivative must be continuous ($C^1$). Choosing 
$C^1$ elements in an arbitrary triangular discretization is not 
trivial, however. 
To overcome this difficulty, we use a mixed variational principle 
\citep{Arnold-1990}, based on the work of Herrmann and Miyoshi 
\citep[see][]{Miyoshi-1976,Reinhart-1982,Blum-Rannacher-1990,Oukit-Pierre-1996},
 which involves treating the moment tensor $M_{\al\be}$ as a separate variable.
This allows us to treat the variables as continuous, and affine over each 
triangular element. For a graphene sheet with clamped conditions at the 
boundary, the six weak form equations that result are (see 
\appref{ap:elasticity-eq-conds} for a detailed derivation):
\numparts
\begin{eqnarray}
\int_\Om\sbra{\pd{\tilde{v}_1}{X}N_{11}+\pd{\tilde{v}_1}{Y}N_{12}+\tilde{v}_1V_
{v_1}[w,v_1,v_2;\la_i]}\mr{d}^2\mb{X} &=0,\label{weakform1}\\
\int_\Om\sbra{\pd{\tilde{v}_2}{X}N_{12}+\pd{\tilde{v}_2}{Y}N_{22}+\tilde{v}
_2V_ {v_2}[w,v_1,v_2;\la_i]}\mr{d}^2\mb{X}&=0,\\
\int_\Om\biggl\{\pd{\tilde{w}}{X}\bra{-\pd{M_{11}}{X}-\pd{M_{12}}{Y}+N_{11}\pd{
w }{X}+N_{12}\pd{w}{Y}} \qquad & \nonumber \\
+\pd{\tilde{w}}{Y}\bra{-\pd{M_{12}}{X}-\pd{M_{22}}{Y}+N_{12}\pd{w}{X}+N_{
22}\pd{w}{Y}} \qquad & \nonumber \\
+\tilde{w}V_w[w,v_1,v_2;\la_i]\biggr\}\mr{d}^2\mb{X}&=0\\
\int_\Om\sbra{\frac{1}{D(1-\nu^2)}(M_{11}-\nu 
M_{22})\tilde{M}_{11}+\pd{w}{X}\pd{\tilde{M}_{11}}{X}}\mr{d}^2\mb{X}&=0,\\
\int_\Om\sbra{\frac{1}{D(1-\nu)}M_{12}\tilde{M}_{12}+\frac{1}{2}\pd{w}{X}
\pd 
{\tilde{M}_{12}}{Y}+\frac{1}{2}\pd{w}{Y}\pd{\tilde{M}_{12}}{X}}\mr{d}^2\mb{X}
&=0 
,\\
\int_\Om\sbra{\frac{1}{D(1-\nu^2)}(M_{22}-\nu 
M_{11})\tilde{M}_{22}+\pd{w}{Y}\pd{\tilde{M}_{22}}{Y}}\mr{d}^2\mb{X}&=0.
\label{weakform6}
\end{eqnarray}
\endnumparts

As will be clear shortly, these equilibrium equations provide the physical 
constraints to the optimization procedure. Our task is to seek a set of 
control parameters (substrate topography, boundary shape, etc.) that, upon 
solution of the variational problem to find the equilibrium configuration of 
the elastic medium, yields a PMF distribution that best approaches the 
prescribed target. The implementation of this optimization is done numerically. 
We have chosen to use piecewise affine finite elements combined with a 
patch recovery method in our calculations for their simplicity and ease of 
implementation (see \appref{ap:strainrecovery}). But it should be noted that 
the method allows higher-order elements to be used, as long as one ensures that 
those formulations are stable and solvable.

\subsection{Coupling deformations to electrons}%

To the weak-form elastic equilibrium equations we must add an equation linking 
the strain field to the generated PMF, $B(\bX)$. This is because we wish to 
find the deformation field that best approximates $B(\bX)$ to a desired target, 
say $\hat{B}(\bX)$. The origin of this PMF that appears in the low-energy 
effective Hamiltonian of deformed graphene is the local modification of the 
electronic hopping amplitudes, $t$, between neighboring carbon atoms brought 
about by the space dependent deformation of the crystal lattice. The hopping is 
constant in the perfect crystal: $t_0=2.7$\,eV. But, since $t$ depends strongly 
on the inter-atomic distance, any local change caused by a deformation leads to 
perturbations to this equilibrium value and, hence, more generically, 
$t\bigl(\bX_i,\bX_i+\bm{n}\bigr)=t_0+\delta t\bigl(\bX_i,\bX_i+\bm{n}\bigr)$. 
The presence of $\delta t$, which is a relatively small perturbation to $t_0$ 
in practical situations, adds a correction to the low-energy Dirac-like 
Hamiltonian that emerges from a tight-binding description of the electronic 
hopping among $p_z$ bands of adjacent carbon atoms. The effective Hamiltonian 
around the point 
$\bm{K}=(4\pi/3\sqrt{3}a,\,0)$ in the first Brillouin zone has the form 
\cite{Kane:1997,Suzuura:2002}
\begin{equation}
  \mathcal{H} = v_F\,
  \bm{\sigma}\cdot\bigl(\bm{p}+e\bm{\Acal}\bigr),
  \label{eq:HDirac}
\end{equation}
where $\bm{\sigma}$ is a vector of Pauli matrices, and $\hbar v_F = 3 t_0 a/2$, 
with $a=1.42$\,\AA~the carbon-carbon distance in equilibrium. For deformations 
on scales that are large compared to $a$, the components of the pseudomagnetic 
vector potential $\bm{\Acal}=\Acal_{x}\bm{e}_x+\Acal_{y}\bm{e}_y$ are 
explicitly given by \cite{Suzuura:2002}
\begin{equation}
  \Acal_x(\bX) - \im \Acal_y(\bX) \simeq -\frac{\hbar c}{2 e a}
  \bigl(\varepsilon_{xx} - \varepsilon_{yy} + 2\im\,\varepsilon_{xy}\bigr)
  ,
  \label{eq:Adef}
\end{equation}
where $c = - \partial\log t(r) / \partial\log r\bigr|_{r=a}$ (see 
\appref{ap:coupling}). For static deformations, a value $c \approx 3.37$ 
captures the changes in various physical properties arising from strain-induced 
modifications of the $\pi$ bands in agreement with first-principles 
calculations \cite{PereiraPRB2009,Ribeiro:2009,Ni:2010,Farjam:2010,Son:2010, 
Pereira-OpticalStrain:2010}. We note, however, that the effective low-energy 
Hamiltonian (\ref{eq:HDirac}) contains only the leading order corrections 
arising from non-uniform deformations; further expanding in higher orders of 
smallness in the strain magnitude and the momentum with respect to $\bm{K}$ 
leads to terms that introduce, for example, Fermi surface anisotropy 
\cite{Hasegawa2006Zero,Pereira-OpticalStrain:2010} and space-dependent $v_F$ 
\cite{deJuan2013Gauge,BarrazaLopez-Discrete,Manes2013Generalized,
RamezaniMasir2013Pseudo}. 
For simplicity, since we want to tailor only the 
PMF distribution as illustration of the method, and to 
keep the focus on the optimization framework rather than the details of the 
different levels of approximation for the effective strain-dependent 
Hamiltonian, we shall focus the subsequent analysis on the Hamiltonian 
(\ref{eq:HDirac}). But it should be clear that, as far as the optimization 
procedure is concerned (which does not take into account the energy of the 
electronic system), the particular form of $\mathcal{H}$ is only relevant in 
order to identify the target quantity that we wish to optimize and its 
expression in terms of the strain components, as in (\ref{eq:Adef}). If instead 
of the PMF we were interested in optimizing, for example, towards a desired 
space-dependence of the Fermi velocity \cite{deJuan2013Gauge} or that of the 
deformation potential \cite{Suzuura:2002}, the method requires only the 
specification of its functional dependence on strain.

Finally, the pseudomagnetic field itself, $B$, being defined as the 2D curl of 
$\bm{\Acal}$, reads: 
\begin{equation}
  B(\bX) = \frac{\hbar c}{ae}
  \sbra{\pd{}{Y}\bra{\frac{\ep_{11}-\ep_{22}}{2}}+\pd{\ep_{12}}{X}}.
  \label{eq:PMF-def}
\end{equation}

As noted above, by virtue of our choice of piecewise affine finite elements for 
the numerical interpolation, the six variables $v_1$, $v_2$, $w$, $M_{11}$, 
$M_{12}$, and $M_{22}$ are treated as continuous, and affine over each 
triangular element.
As a consequence, the strain field will be discontinuous and constant in each 
triangular element, leaving the PMF (\ref{eq:PMF-def}) undefined 
within this interpolation scheme. To overcome this we use the technique of 
patch recovery \cite{Zienkiewicz-Zhu-1992} detailed in 
\appref{ap:strainrecovery}. 
In brief, this is a mechanism that uses the discontinuous strain data 
$\ep_{\al\be}$ to recover a strain field $\ep_{\al\be}^{\mathrm{rec}}$ of the 
same type as the primary variables: continuous and affine over each element. 
The derivative of $\ep_{\al\be}^{\mathrm{rec}}$ is well-defined, and thus so is 
the PMF if it is calculated using this recovered strain field.

\section{Optimization}\label{sect:optimization}

Solution of the weak form variational (equilibrium) equations constitutes 
the \emph{forward problem}:\ in other words, given a set of control variables 
(here chosen to be substrate shape, encoded in the external potential $V$), 
what deformation and PMF do these conditions impose on the graphene sheet? This 
report is aimed at answering the corresponding \emph{inverse problem}: what are 
the control variables that will give rise to a desired PMF?

This inverse question is posed as an optimization problem, where an integral is 
minimized subject to the weak form equations written explicitly in 
Eqs.~(\ref{weakform1}--\ref{weakform6}).
If we let $\hat{B}(X,Y)$ be the desired PMF in Lagrangian coordinates, we then 
seek to minimize the functional
\begin{equation}
\mathcal{I}[w,v_1,v_2]=\int_\Om\bra{B[w,v_1,v_2]-\hat{B}(X,Y)}^2\mr{d}^2\mb{X},
\end{equation}
to find a PMF, $B$, which is (ideally everywhere) as close as possible to 
the prescribed $\hat{B}(X,Y)$ (the reader will note once more at 
this stage that the quantity $B[w,v_1,v_2]$, which is here associated with the 
PMF, can be replaced by any other of interest, as long as its dependence on the 
strain or deformation field can be specified; the scope of applicability of this 
method extends, therefore, well beyond the PMF example chosen here for 
definiteness).

This sort of optimization problem, however, is typically mathematically 
ill-posed, in the sense that there are infinitely-many solutions to such a 
minimization and, in order to find a solution which also satisfies the weak 
form 
equations, the numerical method often yields a solution which is not smooth.
To counter this phenomenon, one must add to the minimization integral, 
$\mathcal{I}$, a regularization term which penalizes high spatial variations in 
the control variables $\la_i$:
\begin{equation}
\mathcal{I}[w,v_1,v_2]=\int_\Om\bra{B[w,v_1,v_2]-\hat{B}(X,Y)}^2\mr{d}^2\mb{X}
+\eta\,\mc{I} ^{\mathrm{
reg}}[\la_i],\label{eq:objfun}
\end{equation}
where $\eta$ is a tunable parameter.
The precise form of $\mc{I}^{\mathrm{reg}}[\la_i]$ will depend on what the 
control variables $\la_i$ represent; for the specific example of substrate 
shape optimization a typical form will be discussed below.
Thus the full problem is to minimize, by varying the six state variables $v_1$, 
$v_2$, $w$, $M_{11}$, $M_{12}$, $M_{22}$ and control variables $\la_i$, the 
objective function (\ref{eq:objfun}) subject to the six equations 
(\ref{weakform1}--\ref{weakform6}), solved for all admissible variations 
$(\tilde{\cdot})$.
In these expressions, $B$ is given by (\ref{eq:PMF-def}), $N_{\al\be}$ is 
given by (\ref{NM-def}), and $\ep_{\al\be}$ is given by (\ref{curvature-vk}).
This problem is an example of a PDE-constrained optimization. For technical 
details regarding well-posedness and solution methods for such problems the 
reader is referred to \citet{Troltzsch-2010} or Borz\`{\i} and Schulz 
\cite{Borzi-Schulz-2011}. This general procedure has also been applied to 
shape optimization in elastic plates experiencing differential growth fields 
\cite{Jones:OptimalShaping}. 

Finally, in solving this problem numerically, all equations are 
adimensionalized in such a way that most variables are $O(1)$ to ensure good 
numerical behavior (details described in \appref{ap:nondimensionalization}).

\section{Practical application: optimizing substrate shapes}

We wish to apply the previously-developed general theory to a specific 
example, to illustrate its practical implementation, and its utility in the 
problem at hand.
The example we have in mind is of a graphene 
sheet forced to conform to a certain substrate shape by pressure and adhesive 
forces, as depicted in \Fref{fig:expt}. The elevation of the substrate is 
denoted $z=\hat{z}(x,y)$ (we use coordinates 
$x$, $y$ as Eulerian coordinates rather than the Lagrangian $X$, $Y$ used in 
the definition of the graphene deformation). For definiteness, we assume that 
we are searching for target PMFs $\hat{B}$ of typical scale 
$B_0=10\,\mathrm{T}$. We further assume that the domain $\Om$, representing the 
shape of the computational domain, has typical dimensions $L=100\,\textrm{\AA}$.

\subsection{The external forces}

As discussed above, the external potential term $V[v_1,v_2,w;\la_i]$ will have 
two components:\ the work done due to hydrostatic pressure, $pw$, and the 
adhesion energy $V_{\mathrm{adh}}$ between the graphene flake and the 
substrate. 
To find the adhesion energy, consider the graphene sheet as a collection of 
atoms, interacting with a field $V_p(x,y,z)$ in $\mathbb{R}^3$-space, such that 
a particle $\mr{d}S$ of the plate, located at $(X+v_1,Y+v_2,w)$ in Eulerian 
coordinates, contributes $V_p(X+v_1,Y+v_2,w)\mr{d}S$ to the adhesion energy.
Then $V_{\mathrm{adh}}[w,v_1,v_2;\la_i]=V_p(X+v_1,Y+v_2,w)$.

Given a substrate shape $z=\hat{z}(x,y)$, we could determine the adhesion 
potential $V_p$ at every point in three-dimensional space. But this will be 
time-intensive in general, and for optimization problems could prove 
prohibitively expensive.
As an alternative, assume that the gradient of the substrate is small, so that 
we can approximate $V_p(x,y,z)=J(z-\hat{z}(x,y))$, where $J(s)$ is some 
one-dimensional adhesion potential, such as the Lennard-Jones potential between 
surfaces \citep{Muller-Yushchenko-Derjaguin-1980}, 
\begin{equation}
J(s)=\frac{J_0}{3}\sbra{\bra{\frac{s^*}{s}}^8-4\bra{\frac{s^*}{s}}^2},
\end{equation}
where $s^*$ is the adhesion well position, or the distance from the substrate 
at which a particle is in equilibrium.
Thus,
\begin{equation}
V[w,v_1,v_2;\la_i]=pw+J\bigl(w-\hat{z}(X+v_1,Y+v_2)\bigr),
\label{eq:interaction-potential}
\end{equation}
with the control variables $\la_i$ encoded in $\hat{z}$.
As a representative value for $s^*$, we use the value that 
\citet{Xu-Buehler-2010} give for C--Cu, namely $2.243\,\textrm{\AA}$.
Similarly we use $J_0=0.45\,\mathrm{J\,m^{-2}}$ as a representative value, from 
the investigation of \citet{Koenig-Boddeti-etal-2011} into the adhesion 
strength between graphene and SiO${}_2$. We select a typical value for the 
hydrostatic pressure as $p=100\,\textrm{bar}=10^7\,\mathrm{Pa}$.

\subsection{Parametrization of the substrate 
topography}\label{sect:substrate-parametrization}

In writing an expression for the substrate geometry, a na\"\i{}ve approach 
would 
be to use the same type of discretization as for the graphene sheet itself. The 
domain $\Om$ is discretized into a collection of triangles, and each of the 
state variables ($w$, $v_\al$, $M_{\al\be}$) is posited to be continuous and 
affine over each triangle. This allows each variable to be described entirely 
in 
terms of its values at the nodes of the triangulation, and is the main 
advantage 
in writing the state equations in weak form
\footnote{In this formulation the variations $\tilde{\cdot}$ are, for each 
nodal 
point $i=1,\ldots,N_p$, the piecewise affine functions which take the value $1$ 
at nodal point $i$ and zero at each other nodal point. For a given weak form 
equation this provides $N_p$ equations for each of the $N_p$ unknown values of 
the function at the nodal points.}.
However, this approach will not work when it comes to describing the shape of 
the substrate, $z=\hat{z}(x,y)$. Since the graphene sheet can move in a lateral 
direction, the triangulations of the substrate and the sheet itself will not 
remain in registration. Therefore, for a given nodal point the distance measure 
$w-\hat{z}(X+\ep v_1,Y+\ep v_2)$ will not vary smoothly as $v_\al$ are varied.
The alternative, which we will follow here, is to construct a smooth shape for 
the substrate.

For the numerical experiments in this article, we assume that the substrate is 
patterned periodically in the two horizontal directions, and set the repeating 
2D unit cell to be a rhombus. If we introduce the two coordinates 
$\xi_1=\frac{y}{\sqrt{3}}+x$ and $\xi_2=\frac{y}{\sqrt{3}}-x$ the unit cell 
corresponds to the domain
\begin{eqnarray}
\Om=\left\{(\xi_1,\xi_2)\;:\;0\leq\xi_1<1,\;0\leq\xi_2<1\right\}.
\end{eqnarray}
The topography of the substrate, $\hat{z}(x,y)$, can then be resolved into a 
truncated Fourier expansion with the period of the unit cell $\Om$. Since the 
(finite) set of expansion coefficients determines the overall topography, they 
play the role of the control variables $\la_i$: varying the topography of the 
substrate is, therefore, achieved by varying these expansion constants (refer 
to \appref{ap:substrate-topography} for particulars of this approach).
The periodicity of the substrate places limits on the patterns of PMF that can 
be sought. If we integrate the PMF over the unit rhombus, we find
\begin{equation}
  \int_\Om B\,\mr{d}^2\mb{X}
  =\oint_{\partial\Om}\sbra{ 
    \bra{\frac{\ep_{11}-\ep_{22}}{2}}n_2+\ep_{12}n_1}\mr{d}s 
  =0,
\end{equation}
because the strain fields generated by the periodic substrate will also be 
periodic. 
Thus (if we limit ourselves to periodic substrates) it is impossible to 
generate PMFs whose integral over the unit cell is nonzero --- in particular 
this rules out the generation of strictly constant nonzero PMFs by periodic 
deformations. It should be emphasized, however, that our general procedure is 
applicable to arbitrary domains, geometries, and target PMFs, which require 
different parametric expansions of the substrate shape in place of the Fourier 
series employed here. The periodic choice is used by us simply out of 
convenience, precisely for its straightforward Fourier expansion that allows the 
description of an arbitrarily patterned substrate using mathematically simple trigonometric functions.

As discussed earlier, in order to avoid convergence towards solutions that are 
ill-behaved during the numerical optimization, a regularization term, 
$\mc{I}^{\mathrm{reg}}$, is added to the objective integral, as per 
\Eqref{eq:objfun} (in this case such ill-behaved solutions could be, for 
example, substrate profiles with discontinuities or sharp topographical 
features). We choose it to be
\begin{equation}
\mathcal{I}^{\mathrm{reg}}=\frac{1}{\mathrm{Area}(\Om)}\int_\Om|\mb{\nabla}
\hat{z}|^2\mathrm{d}^2\mb{X}
\end{equation}
which is simple to calculate using the orthogonality of the basis functions in 
the Fourier expansion over $\Om$. This expression measures the fineness of 
spatial variation in the substrate:\ 
$\mathcal{I}^{\mathrm{reg}}$ is larger for substrate profiles with smaller 
wavelengths. The minimization of (12) thus leads to the penalizing of such 
rough profiles that would be unrealistic in view of the finite 
feature resolution associated with any experimental approach to substrate 
patterning. Moreover, if a particular experimental implementation is to be 
carried out, the regularization term can be further refined or adapted to 
reflect the specific geometric, fabrication, or other constraints.

\begin{figure*}
  \centering
  \includegraphics[width=0.95\textwidth]{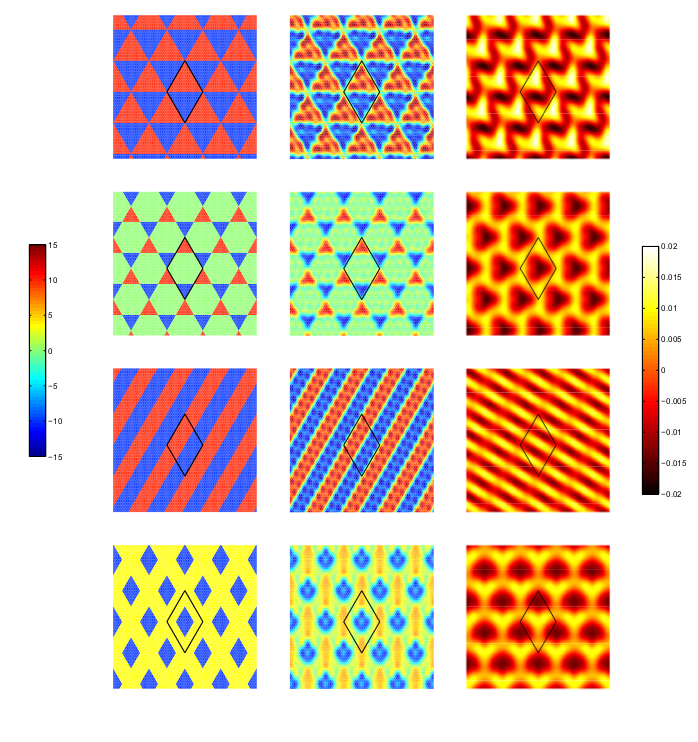}%
  \caption{
    Results of the numerical optimization to make the PMF approach a 
    prescribed spatial pattern with typical magnitude of 10\,T. 
    Column 1: the target PMF patterns (color scale in Tesla, on the left).
    Column 2: calculated PMF (left-hand color scale). 
    Column 3: substrate topography associated with the PMF shown on its left    
 (color scale in units of $L=10\,\mathrm{nm}$, on the right). 
    The unit cell is displayed in each image, and has edges of length $L$.
  }
  \label{fig:B10_results}
\end{figure*}

\begin{figure*}
  \centering
  \includegraphics[width=0.95\textwidth]{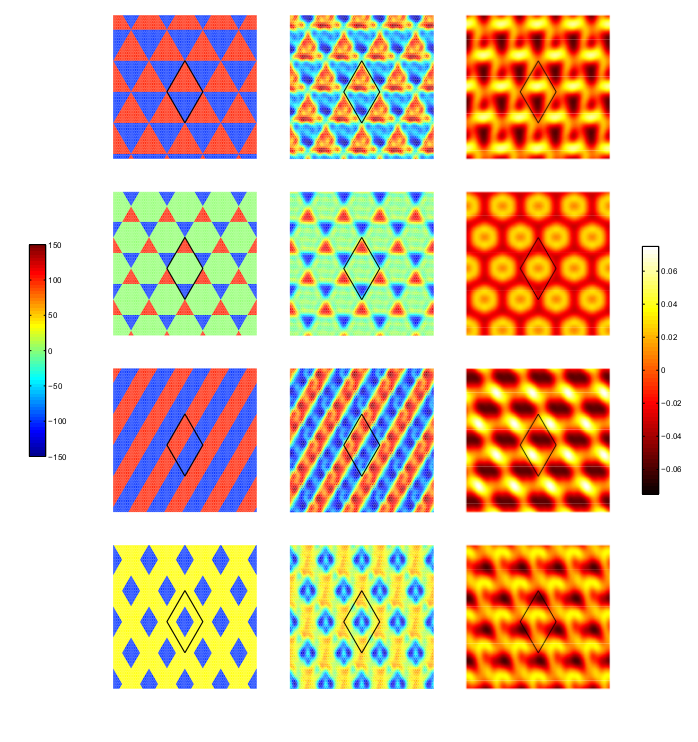}%
  \caption{
    Results of the numerical optimization to make the PMF approach a 
    prescribed spatial pattern with typical magnitude of 100\,T. 
    Column 1: the target PMF patterns (color scale in Tesla, on the left).
    Column 2: calculated PMF (left-hand color scale). 
    Column 3: substrate topography associated with the PMF shown on its left 
    (color scale in units of $L=10\,\mathrm{nm}$, on the right). 
    The unit cell is displayed in each image, and has edges of length $L$.
  }
  \label{fig:B100_results}
\end{figure*}

\section{Results and discussion}

For illustration we initially seek PMFs with  a typical strength of 
$10\,\mathrm{T}$. The unit cell is chosen to have edges of length 
$L=10^{-8}\,\mathrm{m}$ and we apply a pressure of $100\,\mathrm{bar}$. 
These parameters might seem at the threshold of current experimental 
applicability, but they are chosen for their numerical tractability---since if 
$s^*$ is too small, standard numerical algorithms will iterate over trial 
configurations with negative graphene--substrate separations $s$, a highly 
non-smooth problem. To overcome this issue, one would need to carefully design 
algorithms in which negative separations were avoided. But, for the 
proof-of-principle calculations reported here, we choose this acceptable 
compromise.

We obtain good results for the numerical parameters $K=2$, $\eta=10^{-9}$, and 
for an isometric mesh of 800 triangles in the unit cell. For illustration in 
this report we have chosen the following four target PMF patterns to optimize 
for:
\begin{enumerate}
\item An alternating pattern of equilateral triangles,
\item A kagome-like pattern of zero-valued hexagons interspersed with 
alternating triangular regions,
\item Alternating strips of positive and negative fields,
\item A rhombus of negative PMF surrounded by a field of positive PMF.
\end{enumerate}
Each of these target fields (shown in the first column of 
\Fref{fig:B10_results}) adheres to the condition that its integral over the 
unit cell must be zero. In the second column we see the PMF attained by the 
optimization code, and in the third column we see the substrate topography that 
produced such a field.\footnote{
Three-dimensional plots of the optimized graphene topographies due to the substrate shapes in Figures \ref{fig:B10_results} and \ref{fig:B100_results}, together with the resulting PMFs, are shown in \ref{app:3dplots}.
} %
It is clear that the converged solutions reproduce with 
very good accuracy the spatial dependence of the target PMF in all four cases, 
including the rapid sign changes imposed by the target field which, being 
necessarily smooth in the solution, are still quite sharp, with the sign change 
occurring in a very short length scale. This documents how this optimization 
strategy is able to capture all the features, global and detailed, of the 
target 
PMF. From an experimental point of view, the power of this method is clear: by 
providing an accurate solution to the inverse problem, it allows one to specify 
in detail what substrate pattern and topography leads to a PMF of given 
magnitude and space dependence. This, in principle, provides all the 
experimental information needed to fabricate the corresponding structures.

It is important to recall that, as pointed out earlier, the solution to the 
pseudomagnetic inverse problem might not be unique. This is because the 
objective function (\ref{eq:objfun}) is defined in terms of the field $B$, 
whose 
relation to the deformation field expressed in \Eqref{eq:PMF-def} allows for a 
large ``gauge'' freedom. As a result, more than one set of control parameters 
within a certain parameter range might be simultaneously compatible with the 
target field within a desired accuracy and obey the elastic plate equilibrium 
equations. On the other hand, having found a set of parameters that optimizes 
the induced PMF to the target sought is not a guarantee that such set will 
remain optimal upon finite changes of an external variable, or a constant 
scaling of the target function. This latter aspect is best illustrated with a 
specific case for our example system. The panels in the last column of 
\Fref{fig:B10_results} show the suitable substrate profiles for a pressure of 
100\,bar and PMFs with an amplitude of $10$\,T. The equivalent calculations for 
a target PMF magnitude of $100$\,T, a ten-fold increase, yield the results 
shown 
in \Fref{fig:B100_results}. It is clear that the optimal substrate topographies 
that guarantee the same degree of proximity between the induced and target PMF 
as before are markedly different from those in \Fref{fig:B10_results}. This 
means that in some experimental setups, such as the one sketched in 
\Fref{fig:expt}, the proximity of the induced PMF to the target might need to 
be compromised in favor of a having a fixed set of control parameters suitable 
in a range of PMF field amplitudes (i.e., a single substrate profile able to 
generate acceptable PMF profiles of different amplitude). However, the power of 
the method and its experimental practicality in allowing a direct, one-step 
route from PMF design to substrate fabrication, should largely compensate such 
compromises, when they are unavoidable.

There are many ways in which one can experimentally control the deformation of 
a graphene sheet \cite{Lu-Bao-etal-2013}, with each choice leading to a 
different set of control variables. Most obviously, the graphene sheet may be 
manipulated directly, whether by substrate topology (such as nanopillars) 
\cite{NeekAmal-Covaci-Peeters-2012,Tomori-Kanda-etal-2011}, a distribution of 
attached structures like nanotubes \cite{Lim-Koon-etal-2012}, or a 
nano-manipulation of substrate adhesion properties \cite{LuNatureComm2012}. The 
corresponding control variables would be the configurations of the 
nanostructures, including their shapes, their positions in relation to the 
graphene sheet, and their height.
Edge actuation, where the control variables are the displacements applied to 
the 
edge, is another deformation mode 
\cite{Guinea-Geim-etal-2010,Guinea-Katsnelson-Geim-2009}. In experiments, it 
may 
be preferable to apply these edge displacements indirectly, by applying 
electromechanical forces to the electrodes attached to a graphene flake 
\cite{Zhang-Huang-etal-2014}. The position and shape of the electrodes form the 
control variables in this case.

A further class of deformations in graphene are the inflation of bubbles by 
suspending graphene over a particularly-shaped cavity, whether by hydrostatic 
pressure or electromechanical forces 
\cite{Georgiou-Britnell-etal-2011,Klimov-Jung-etal-2012}. Since the forcing in 
these examples is global, it is the shape of the cavity that provides the 
variation in the calculated strain field, and as such the control variables in 
an optimization calculation would be a parametrization of the cavity shape. 
Such 
inflation problems can be coupled with local deformation, in the form of a 
point 
deformation due to a STM tip \cite{Xu-Yang-etal-2012,Klimov-Jung-etal-2012}, 
providing additional control variables of tip position and strength and 
allowing 
a greater ability to achieve desired strain fields and consequential electronic 
properties.

Finally, it should be noted that, despite our focus in this report on 
optimizing the control parameters for a target PMF (which constitutes the core 
of the strain-engineering concept in graphene), this optimization framework can 
be rather easily extended to other target quantities by replacing the 
objective function in \Eqref{eq:objfun} by the relevant measure of 
``distance'' for that problem, and specifying its dependence on the strain or 
displacement field analogously to the specification in \Eqref{eq:PMF-def}. 
Each of the three main components of the procedure---namely the \emph{objective 
function} that is minimized, the \emph{state equations} that form the 
constraints (here being the elastic plate equations), and the \emph{control 
variables} that are open to experimental variation---can be changed to answer 
different questions of interest.

As simple examples, we suggest that one may wish to minimize or maximize the 
degree of rippling obtained in the edge actuation of a suspended graphene sheet 
\cite{Bao-Miao-etal-2009}, or that the resonant frequency of a graphene flake 
suspended over a cavity 
\cite{Bunch-vanderZande-etal-2007,GarciaSanchez-vanderZande-etal-2008} may be 
optimized by varying the cavity shape according to the principles outlined in 
this exposition.

\section{Summary}

We presented a general-purpose framework suitable to answer the following 
inverse problem in graphene: which set of external control parameters 
(substrate topography, sample shape, load distribution, etc.) guarantees that 
the resulting equilibrium state of graphene exhibits a pseudomagnetic field 
that varies in space in a prescribed way? The ability to answer this question 
in 
general, given only a potential experimental setup and the target field 
profile, 
is paramount towards fulfilling the vision of tailored transport and other 
electronic properties in graphene by strain-engineering. This concept calls for 
expedited ways to answer the question above. The method presented here relies 
on 
a PDE-constrained optimization strategy to minimize the generic objective 
function (\ref{eq:objfun}) which penalizes significant deviations between the 
induced and target PMFs. It thus affords a one-step route from PMF design to 
experimental implementation, is unbiased and general enough to accommodate a 
multitude of experimental parameters and conditions that can be envisaged to 
produce the desired deformations in the graphene lattice, and 
always ensures compliance with the constraints imposed by elasticity theory and 
the equilibrium conditions of graphene treated as a continuous elastic medium.
We trust that it can be an important tool in designing or guiding 
experimentally 
realistic conditions for strain-engineered graphene devices and beyond---the 
versatility to define, in principle, any target function for other physical 
quantities entails a broad applicability.

\section*{Acknowledgements}

The authors acknowledge support from the NRF-CRP grant ``Novel 2D materials 
with tailored properties: beyond graphene'' (R-144-000-295-281), as well as 
productive discussions with A.~H.~Castro Neto and L.\ Mahadevan.

% ------------------------------------------------------------------------------
% SUPPLEMENTARY MATERIAL
% ------------------------------------------------------------------------------

\clearpage

% === Change labels and tags for supplementary info:

% \renewcommand{\thesection}{S-\Roman{section}}
% \renewcommand{\theequation}{S-\arabic{equation}}
% \renewcommand{\thepage}{S-\arabic{page}}
% \renewcommand{\thefigure}{Supplementary~Figure~\arabic{figure}}
% \makeatletter 
%   \renewcommand{\figurename}{\hspace*{-0.4em}}
%   \renewcommand{\fnum@figure}{\figurename~\thefigure}
% \makeatother
% % change citation number/label style at the end of the document
% \renewcommand{\bibnumfmt}[1]{[S-#1]}
% % \renewcommand{\bibnumfmt}[1]{\textsuperscript{S-#1}}
% % change citation number/label style within the document
% \renewcommand{\citenumfont}[1]{S-#1}
% 
% % === Switch to onecolumn and shrink the text width
% 
% % \onecolumngrid
% \setlength{\hoffset}{-1in}
% \setlength{\textwidth}{14cm}
% \setlength{\oddsidemargin}{\paperwidth*\real{0.5} - \textwidth*\real{0.5} }
% 
% % === Supplementary cover page:
% 
% \clearpage\hrule
% \section*{\large Supplementary Information}
% \hrule\vspace*{3em}
% 
% \begin{center}
% %   \makeatletter\textbf{\large\@title}\makeatother \par\medskip
%   Gareth W. Jones and Vitor M. Pereira \par\medskip
%   \emph{Graphene Research Centre \& Department of Physics. \\
%   National University of Singapore, 2 Science Drive 3, Singapore 117542}
% \end{center}
% 
% \bigskip\hrule\bigskip

% === Reset counters:

% \setcounter{section}{0}
% \setcounter{subsection}{0}
% \setcounter{subsubsection}{0}
% \setcounter{equation}{0}
% \setcounter{figure}{0}
% \setcounter{page}{1}

\appendix

% \begin{bibunit}[unsrt]

\section{Graphene as an elastic continuum}\label{ap:elasticity}

Here we recapitulate the results of sections \ref{sect:elastic-basics} and 
\ref{sect:elastic-eqm} with further discussion on their validity and 
applicability.

\subsection{Definitions and assumptions}\label{ap:elasticity-defs}

The deformation of a graphene sheet is described in terms of its deviation from 
a flat two-dimensional surface. 
A point $\bX$ in an undeformed flat surface is defined by its coordinates 
$(X_1,X_2,0)=(X,Y,0)$, for all $(X,Y)$ belonging to some set $\Omega$ that 
defines the physical domain. Under a deformation the point $\bX=(X,Y,0)$ is 
transformed to $\br=(x_1,x_2,x_3)=(x,y,z)$ in three-dimensional space, where
\begin{eqnarray}
  x&=X+v_1(X,Y), \nonumber \\ 
  y&=Y+v_2(X,Y), \nonumber \\ 
  z&=w(X,Y),
\end{eqnarray}
are the expressions for the new coordinates in terms of the 
in-plane displacements $v_\alpha$, and the vertical deflection $w$
\footnote{Our convention is to have all Greek indices $\in\{1,2\}$. The 
coordinate system of the undeformed sheet is chosen to be Cartesian, and we 
will make extensive use of Einstein's summation convention throughout this 
report. Subscripts following a comma denote partial differentiation with 
respect to that coordinate.}%
.
From these expressions one can define the base vectors 
$\mb{r}_{,\al}=(x_{,\al},y_{,\al},z_{,\al})$ and the metric tensor 
$g_{\al\be}=\mb{r}_{,\al}\cdot\mb{r}_{,\be}$ of the deformed surface. Then the 
true strain tensor is defined in terms of the difference of this metric tensor 
from its original value $\de_{\al\be}$ (Kronecker's delta, or the identity 
tensor, from our choice of Cartesian coordinates):
\begin{eqnarray}
  \ep_{\al\be}^{\mathrm{true}}=\frac{1}{2}(g_{\al\be}-\de_{\al\be}).
\end{eqnarray}
In terms of displacements, this becomes
\begin{eqnarray}
  \ep_{\al\be}^{\mathrm{true}} = \frac{1}{2}\bra{v_{\al,\be} + v_{\be,\al} + 
    w_{,\al}w_{,\be} + v_{\ga,\al}v_{\ga,\be}} 
  .
  \label{ap-strain-full} 
\end{eqnarray}
The second deformation measure of the surface is the curvature tensor, defined 
by $\rho_{\al\be}=\mb{r}_{,\al\be}\cdot\mb{n}$, where $\mb{n}$ is the unit 
normal vector to the deformed surface.
We do not derive the result here (for details in the case of a curved elastic 
shell see \citet{Koiter-1966} or \citet{Niordson-1985}) but the full expression 
for the curvature tensor in Cartesian coordinates is
\begin{eqnarray}
  \rho_{\al\be}^{\mathrm{true}} = 
  g^{-1/2} w_{,\al\be}\bra{1+v_{\la,\la}+\frac{1}{2}v_{\la,\la}v_{\mu,\mu}
  -\frac{1}{2}v_{\la,\mu}v_{\mu,\la}} \nonumber\\
  -g^{-1/2} v_{\ka,\al\be}w_{,\la} 
    \bra{\de_{\ka\la}+\de_{\ka\la}v_{\mu,\mu}-v_{\la, \ka}}
  ,
  \label{ap-curvature-full}
\end{eqnarray}
where $g=\det g_{\al\be}$.

Expressions (\ref{ap-strain-full}) and (\ref{ap-curvature-full}) are far too 
unwieldy for most purposes.
Based on assumptions regarding the relative sizes of the displacement 
components 
and the length scale of deformations, the strain and curvature tensors are 
simplified. We chose the von K\'arm\'an 
approximation for its simplicity and its capacity to model moderate deflections. 
This simplification uses the expressions
%
% \numparts\label{ap-strain-curvature-vk}
\begin{eqnarray}
  \ep_{\al\be}=\frac{1}{2}\bra{v_{\al,\be}+v_{\be,\al}+w_{,\al}w_{,\be}},
  \label{ap-strain-vk}\\
  \rho_{\al\be}=w_{,\al\be} .
  \label{ap-curvature-vk}
\end{eqnarray}
% \endnumparts
%
Denote the corrections to these approximations by 
$\ep_{\al\be}^{\mathrm{corr}}=\ep_{\al\be}^{\mathrm{true}}-\ep_{\al\be}$, and 
similarly for $\rho_{\al\be}^{\mathrm{corr}}$.
As an \emph{a posteriori} check on the validity of our solutions, we can verify 
that the approximations are close to the true values, or
\begin{eqnarray}
  \ep_{\al\be}^{\mathrm{corr}}\ll\ep_{\al\be},
  \qquad
  \rho_{\al\be}^{\mathrm{corr}}\ll\rho_{\al\be}.
\end{eqnarray}

The stress and moment resultants are assumed to be isotropic and linear in the 
strain and curvature tensors: %
\begin{eqnarray}
N_{\al\be}=C\,\mathbb{A}_{\al\be\ga\de} \; \ep_{\ga\de},\qquad 
M_{\al\be}=D\,\mathbb{B}_{\al\be\ga\de} \; \rh_{\ga\de},\label{ap-NM-def}
\end{eqnarray}
where
%
% \numparts
\begin{eqnarray}
\mathbb{A}_{\al\be\ga\de}=\bra{\frac{1-\nu}{2}}(\de_{\al\ga}\de_{\be\de}+\de_{
\al\de}\de_{\be\ga}) +\nu\de_{\al\be}\de_{\ga\de},\\
\mathbb{B}_{\al\be\ga\de}=\bra{\frac{1-\sigma}{2}}(\de_{\al\ga}\de_{\be\de}+\de_
{\al\de}\de_{\be\ga}) +\sigma\de_{\al\be}\de_{\ga\de},
\end{eqnarray}
% \endnumparts
%
$\nu$ being the Poisson ratio and $C$, $D$ the stretching and bending moduli, 
respectively.
We have defined $\sigma$ to be the analog of the Poisson ratio for bending 
deformations; if $D_G$ is the Gaussian bending rigidity in the Helfrich free 
energy for the bending of a membrane \citep{Helfrich-1973}, then 
$\sigma=1+D_G/D$.
To calculate the stretching modulus $C$, we use the results of 
\citet{Wei-Fragneaud-etal-2009}, who fitted a polynomial stress--strain 
relation 
to \emph{ab initio} calculations up to strains of $50\%$. For simplicity, we 
will assume a linear stress--strain relationship, which is valid only up to 
strains of around $10\%$.
The linear terms of \citet{Wei-Fragneaud-etal-2009} are, in our notation,
\begin{equation}
  \left.
  \begin{array}{l}
    C=358.1\,\mathrm{N\,m^{-1}} \\
    C\nu=60.4\,\mathrm{N\,m^{-1}}
  \end{array}
  \right\}
  \quad\Rightarrow\quad\nu=0.169.
\end{equation}
This value of $\nu$ agrees with the experimentally-measured Poisson ratio in 
graphite \cite{Blakslee:1970}. The value we chose for the bending modulus was 
that of \citet{Kudin-Scuseria-Yakobson-2001}, calculated \emph{ab initio} as 
$D=1.46\,\textrm{eV}=2.34\times10^{-19}\,\textrm{N\,m}$. We have found only two 
investigations into the value of $D_G$ (and hence $\sigma$) in graphene; the 
calculations of \citet{Wei-Wang-etal-2013} lead to $\sigma=-0.056$, whereas the 
numerical study of \citet{Koskinen-Kit-2010} gives a value of $\sigma=0.565$. 
In 
the absence of consensus, in our calculations $\sigma$ is set to be equal to 
the 
Poisson ratio: $\sigma=\nu=0.169$, and thus 
$\mathbb{B}_{\al\be\ga\de}=\mathbb{A}_{\al\be\ga\de}$. 

The constitutive equations for macroscopic materials are usually derived from 
full three-dimensional isotropic elasticity expressions in the limit that the 
plate thickness is small. In the most rigorous treatments this analysis leads 
to 
limits on the validity of simplifying expressions such as 
(\ref{ap-curvature-vk}) in terms of the relative sizes of stored elastic 
energy, applied surface tractions, and plate thickness 
\citep{Ciarlet-1980,Ciarlet-1997,Friesecke-James-Muller-2006}.
Using such analyses, $\sigma=\nu$ and the stiffness and bending moduli may be 
written in terms of the three-dimensional Young's modulus $E$ and the thickness 
$h$:
\begin{eqnarray}
  C=\frac{Eh}{1-\nu^2},\qquad D=\frac{Eh^3}{12(1-\nu^2)}.
\end{eqnarray}
Using a typical value $h\approx0.3\,\textrm{nm}$ for graphene thickness 
\cite{Huang:2006}, this has been used to cite graphene's Young's modulus as of 
the order of $1\,\textrm{TPa}$ \cite{Lu:1997,Hernandez:1998,LeeSCIENCE2008}. 
While this may be useful to convey the scale and exceptional strength of 
graphene, the same numbers lead to an inaccurate value for the bending modulus 
$D$. Treating graphene as a continuous 3D elastic object is a convenient 
approximation, so for definiteness we keep the two-dimensional 
parameters $C$ and $D$ as our main quantities here rather than express them in 
terms of Young's modulus $E$. A rigorous justification of the plate equations 
used to model graphene deformations is beyond the scope of this paper, and 
would 
involve a detailed analysis of the stored energy involved together with applied 
surface tractions. For the purposes of this paper it is enough to ensure that 
the strains and curvatures are within reasonable limits ($|\ep_{\al\be}|<0.1$, 
$|\rho_{\al\be}|<h^{-1}$) and that the corrections to the strains and 
curvatures 
are small.

\subsection{Weak form equations}\label{ap:elasticity-eq-conds}

In section \ref{sect:elastic-eqm} we stated that equations 
(\ref{weakform1}--\ref{weakform6}) 
could be derived from a minimization of the energy integrals 
(\ref{eq:energy1}--\ref{eq:energy2}). 
In this section we justify this claim.

Recall that the stored energy in the plate was given by
\begin{equation}
\mc{E}=\int_{\Om}\bra{\frac{1}{2}\ep_{\al\be}N_{\al\be}+\frac{1}{2}\rh_{\al\be}
M_{\al\be}+ V[w,v_1,v_2;\la_i]}\mr{d}^2\mb{X}.\label{ap-eq:appenergy}
\end{equation}
To help understand the mixed variational principles that we rely on, let us 
consider a simplified problem of purely transverse deflections of an elastic 
plate subject to  hydrostatic pressure:
\begin{equation}
\hspace*{-3em}
\mc{E}^\mathrm{t}=\int_{\Om}\bra{\frac{1}{2}\rh_{\al\be}M_{\al\be}
+ 
pw}\mr{d}^2\mb{X}=\int_{\Om}\bra{\frac{D}{2}\mathbb{B}_{\al\be\ga\de}w_{,\al\be} 
w_{,\ga\de}+ pw}\mr{d}^2\mb{X}.\label{ap-eq:transverse-plate-energy}
\end{equation}
Assume that the boundary of this plate is formed of three disjoint regions: 
$\partial \Om=\Gamma_c\cup\Gamma_s\cup\Gamma_f$, with clamped conditions along 
$\Gamma_c$, simply supported conditions along $\Gamma_s$, and free conditions 
along $\Gamma_f$.

The standard variational approach is to minimize $\mathcal{E}^\mathrm{t}$ over 
the space of all twice-differentiable $w$ satisfying $w=0$ on 
$\Gamma_c\cup\Gamma_s$ and $\partial_n w=0$ on $\Gamma_c$. The first variation 
of $\mathcal{E}^\mathrm{t}$ is
\begin{equation}
\delta\mathcal{E}^\mathrm{t}=\int_\Om\bra{D\mathbb{B}_{\al\be\ga\de}w_{,\al\be}
\tilde{w}_{,\ga\de}+p\tilde{w}}\mathrm{d}^2\mb{X},\label{ap-eq:weakform_simple}
\end{equation}
where $\tilde{w}=\delta w$ is the variation in $w$.
The weak solution is then the twice-differentiable function $w(X,Y)$ that 
satisfies $\delta\mathcal{E}^\mathrm{t}=0$ for each twice differentiable 
variation $\tilde{w}$ satisfying $\tilde{w}=0$ on $\Gamma_c\cup\Gamma_s$ and 
$\partial_n\tilde{w}=0$ on $\Gamma_c$.
To find the strong form equation and boundary conditions to which this weak 
formulation corresponds, assume that $w$ is four-times differentiable and 
integrate (\ref{ap-eq:weakform_simple}) twice by parts:
\begin{eqnarray}
\fl \de\mathcal{E}^\mathrm{t} = 
  \int_\Om\tilde{w}
    \bra{D\mathbb{B}_{\al\be\ga\de}w_{,\al\be\ga\de}+p}
    \mathrm{d}^2\mb{X} 
  +\oint_{\partial\Om}
  \bra{\tilde{w}_{,\al}M_{\al\be}n_{\be}-\tilde{w}M_{\al\be,\al}n_\be}
  \mathrm{d}s \nonumber\\
\fl = \int_\Om\tilde{w}
  \bra{D\mathbb{B}_{\al\be\ga\de}w_{,\al\be\ga\de}+p}
  \mathrm{d}^2\mb{X} 
  +\oint_{\partial\Om}
  \Bigl[(\partial_n\tilde{w}n_\al + \partial_t\tilde{w}t_\al) 
  M_{\al\be}n_{\be} -   \tilde{w}M_{\al\be,\al}n_\be \Bigr]
  \mathrm{d}s \nonumber\\
\fl=\int_\Om\tilde{w}
  \bra{D\mathbb{B}_{\al\be\ga\de}w_{,\al\be\ga\de}+p}\mathrm{d}^2\mb{X} 
  + \int_{\Gamma_s\cup\Gamma_f}\pd{\tilde{w}}{n}M_{\al\be}n_\al 
  n_\be\,\mathrm{d}s \nonumber \\
-\int_{\Gamma_f}\tilde{w}
    \sbra{\pd{}{s}(M_{\al\be}t_\al n_\be)+M_{\al\be,\al}n_\be}
    \mathrm{d}s,
\end{eqnarray}
using the boundary conditions for $\tilde{w}$.
Setting this to zero for each admissible variation $\tilde{w}$, we find that 
the 
governing equation is
\begin{equation}
D\nabla^4w+p=0\label{ap-eq:strongform},
\end{equation}
with boundary conditions
%
% \numparts
\begin{eqnarray}
w=\pd{w}{n}=0 \qquad\textrm{on }\Gamma_c,\\
w=M_{\al\be}n_\al n_\be=0 \qquad\textrm{on }\Gamma_s,\\
M_{\al\be}n_\al n_\be=\pd{}{s}(M_{\al\be}t_\al n_\be)+M_{\al\be,\al}n_\be=0 
\qquad\textrm{on }\Gamma_f.
\end{eqnarray}
% \endnumparts
%
This is the Euler--Lagrange equation associated with the minimization of 
(\ref{ap-eq:transverse-plate-energy}).

\subsubsection{Mixed variational principles}\label{ap:mixedvariational}

In a standard variational principle, the weak form equations are found by 
minimizing the energy functional.
In a typical mixed variational principle, a dual variable is selected and a new 
functional is introduced.
For the simple plate bending problem above the dual variable is usually 
selected 
to be the bending moment tensor $M_{\al\be}$ %
\footnote{Though \citet{Reinhart-1982} and others have used the curvature 
tensor 
$\rho_{\al\be}$ in place of $M_{\al\be}$, this merely results in a 
rearrangement 
of the governing equations, since one is a linear combination of the components 
of the other.}. %
The variational functional is a version of the Hellinger--Reissner principle 
\citep{Arnold-1990,Blum-Rannacher-1990}, given by

\begin{equation}
\mc{H}[w,M_{\al\be}]=\int_\Om\sbra{-\frac{1}{2D}\mathbb{E}_{\al\be\ga\de}M_{
\al\be}M_{\ga\de}+w_{,\al\be}M_{\al\be}+pw}\mr{d}^2\mb{X}.\label{ap-eq:HR}
\end{equation}
Here
\begin{eqnarray}
\mathbb{E}_{\al\be\ga\de}=\frac{1}{(1-\sigma^2)}\sbra{\frac{(1+\sigma)}{2}(\de_{
\al\ga}\de_{\be\de}+\de_{\al\de}\de_{\be\ga}) -\sigma\de_{\al\be}\de_{\ga\de}}
\end{eqnarray}
is the inverse of $\mathbb{B}_{\al\be\ga\de}$.
The weak form equations are derived from this principle by finding the 
stationary value of $\mc{H}$ over all functions $M_{\al\be}$ and $w$ satisfying 
the aforementioned conditions on $\Gamma_c$ and $\Gamma_s$. 
Note that this stationary value of $\mc{H}$ will be neither a minimum nor a 
maximum; it is for this reason that these methods are often called saddlepoint 
methods. The weak form equations are found by setting the variation $\de\mc{H}$ 
to zero, where
\begin{equation}
\de\mc{H}=\int_{\Om}\sbra{-\frac{1}{D}\mathbb{E}_{\al\be\ga\de}\tilde{M}_{\al\be
}M_{\ga\de}+w_{,\al\be}\tilde{M}_{\al\be}+\tilde{w}_{,\al\be}M_{\al\be}+p\tilde{
w}}\mathrm{d}^2\mb{X}.
\end{equation}
Performing integration by parts one may once more recover the strong 
formulation (\ref{ap-eq:strongform}) with the correct boundary conditions.

However, the formulation (\ref{ap-eq:HR}) still requires a certain regularity 
of the deflection $\tilde{w}$; broadly speaking the square of its second 
derivative must be integrable.
Meanwhile, the only regularity required of $M_{\al\be}$ is that its square must 
be integrable.
One of the main advantages of the mixed variational approach is that it allows 
regularity to be transferred from the displacement to the moment.
On integrating (\ref{ap-eq:HR}) by parts we obtain
\begin{eqnarray}
\fl \mc{H}[w,M_{\al\be}]=\int_\Om\sbra{-\frac{1}{2D}\mathbb{E}_{\al\be\ga\de}M_{
\al\be}M_{\ga\de}-w_{,\al}M_{\al\be,\be}+pw}\mr{d}^2\mb{X}
+\oint_{\partial\Om} M_{\al\be}w_{,\al}n_\be\,\mr{d}s.\nonumber
\\
\label{ap-eq:HR-ibp}
\end{eqnarray}
This functional is one that can be minimized over the space of all $w$ and 
$M_{\al\be}$ whose first derivatives are square-integrable.
However, such $w$ are unable to account for zero normal-derivatives on the 
boundary, so we encode that information directly in (\ref{ap-eq:HR-ibp}):\ 
since 
$w=w_n=0$ on $\Gamma_c$, the boundary integral in (\ref{ap-eq:HR-ibp}) is zero 
over 
$\Gamma_c$, and hence
\begin{eqnarray}
\fl \mc{H}[w,M_{\al\be}]=\int_\Om\sbra{-\frac{1}{2D}\mathbb{E}_{\al\be\ga\de}M_{
\al\be}M_{\ga\de}-w_{,\al}M_{\al\be,\be}+pw}\mr{d}^2\mb{X} \nonumber 
+\int_{\Gamma_s\cup\Gamma_f}M_{\al\be}w_{,\al}n_\be\,\mr{d}s
.\\
\label{ap-eq:HR-C1}
\end{eqnarray}
The weak form equations are obtained by finding the stationary value of 
(\ref{ap-eq:HR-C1}) over the space of admissible functions satisfying $w=0$ on 
$\Gamma_s\cup\Gamma_f$; in other words
\begin{eqnarray}
\int_{\Om}\sbra{-\frac{1}{D}\mathbb{E}_{\al\be\ga\de}\tilde{M}_{\al\be}
M_{
  \ga\de}-\tilde{w}_{,\al}M_{\al\be,\be}-w_{,\al}\tilde{M}_{\al\be,\be}
  +p\tilde{w}}\mr{d}^2\mb{X} \nonumber \\
+\int_{\Gamma_s\cup\Gamma_f}(\tilde{w}_{,\al}M_{\al\be}+w_{,\al} 
  \tilde{M}_{\al\be})n_\be\,\mr{d}s \nonumber \\
=0
\end{eqnarray}
for all trial functions $\tilde{w}$ and $\tilde{M}_{\al\be}$ satisfying these 
conditions.
Again, these weak form equations lead to the same strong form 
(\ref{ap-eq:strongform}) together with appropriate boundary conditions.

This elementary exposition has omitted technical details regarding the 
regularity of the solutions; for a more rigorous consideration the reader is 
referred to \citet{Arnold-1990}, \citet{Blum-Rannacher-1990}, and 
\citet{Oukit-Pierre-1996}.

\subsubsection{Application to nonlinear plate bending}\label{ap:application}

The application of these mixed variational principles to nonlinear plates was 
first analyzed by \citet{Miyoshi-1976} and \citet{Reinhart-1982}.
They were interested in developing numerical methods to study the buckling of 
compressed plates.
This meant that their boundary conditions were ones of applied force, which 
allowed them to use an Airy stress function approach, leading to coupled 
fourth-order differential equations. 
We are unable to use these equations directly as our boundary conditions are 
ones of zero-displacement, which is difficult to express in terms of the stress 
function. 
Instead, we simply add the in-plane stored elastic energy to the variational 
formulation (\ref{ap-eq:HR-C1}), together with an arbitrary external potential.
Writing $N_{\al\be}=C\mathbb{A}_{\al\be\ga\de}\ep_{\ga\de}$ for simplicity, 
where 
\begin{equation}
\ep_{\al\be}=\frac{1}{2}\bra{v_{\al,\be}+v_{\be,\al}+w_{,\al}w_{,\be}},
\end{equation}
the mixed variational functional we use is
\begin{eqnarray}
\fl \mc{H}[v_\ka,w,M_{\ka\la}] =
  \int_{\Om}\bra{
    \frac{1}{2}N_{\al\be}\ep_{\al\be}
    -\frac{1}{2D}\mathbb{E}_{\al\be\ga\de}M_{\al\be}M_{\ga\de}}\mr{d}^2\mb{X} 
    \nonumber \\
  + \int_{\Om}\Bigl(
    -w_{,\al}M_{\al\be,\be}+V[v_\ka,w] \Bigr)\mr{d}^2\mb{X} 
    +\int_{\Gamma_s\cup\Gamma_f}w_{,\al}M_{\al\be}n_\be\,\mr{d}s.
\end{eqnarray}
According to the discussion above, in order to derive the weak form equations 
we should find the stationary value of $\mc{H}$ over all admissible $v_\al$, 
$w$, $M_{\al\be}$ satisfying $v_\al=w=0$ on $\Gamma_c\cup\Gamma_s$.
The first variation of $\mathcal{H}$ can be straightforwardly derived along the 
same lines discussed above, whereupon one obtains
\begin{eqnarray}
\fl \delta\mc{H}=\int_\Om\bra{\tilde{v}_{\al,\be}N_{\al\be}+\tilde{v}_\al 
V_{v_\al}}\mr{d}^2\mb{X} \nonumber \\
+\int_\Om\bra{-\tilde{M}_{\al\be}\frac{1}{D}\mathbb{E}_{\al\be\ga\de}M_{\ga\de
}-\tilde{M}_{\al\be,\be}w_{,\al}}\mr{d}^2\mb{X}+\int_{\Gamma_s\cup\Gamma_f}
\tilde{M}_{\al\be}w_{,\al}n_\be\,\mr{d}s \nonumber \\
+\int_\Om\sbra{\tilde{w}_{,\al}(N_{\al\be}w_{,\be}-M_{\al\be,\be})+\tilde{w}
V_w}\mr{d}^2\mb{X}+\int_{\Gamma_s\cup\Gamma_f}\tilde{w}_{,\al}M_{\al\be}n_\be\,
\mr{d}s.
\end{eqnarray}
These equations then lead naturally to the weak form equations 
(\ref{weakform1}--\ref{weakform6}) on assuming that the entire boundary is 
clamped ($\Gamma_s=\Gamma_f=\varnothing$), that $\sigma=\nu$, and on writing 
out the equations for the 
six components $\tilde{w}$, $\tilde{v}_\al$ and $\tilde{M}_{\al\be}$ 
explicitly. 

The equations hold for all continuous integrable variations 
  $\tilde{v}_{\al}$, $\tilde{w}$, $\tilde{M}_{\al\be}$ that satisfy 
$\tilde{v}_{\al}=\tilde{w}=0$ on the boundary.

\section{Coupling deformations to electrons: PMF\lowercase{s}}%
\label{ap:coupling}

To the six weak-form elastic equations derived in 
\ref{ap:elasticity-eq-conds} we must add an equation linking the strain field to 
the generated PMF, $B(\bX)$. This is because we wish to find the deformation 
field that best approximates $B(\bX)$ to a desired (target), $\hat{B}(\bX)$. The 
origin of this PMF that appears in the low-energy effective Hamiltonian of 
deformed graphene is the local modification of the electronic hopping amplitudes 
between neighboring carbon atoms brought about by the space dependent 
deformation of the crystal lattice.

A single orbital nearest-neighbor tight-binding model for the $\pi$ bands 
derived from electronic hopping among $p_z$ orbitals of neighboring carbons has 
been extremely successful in describing the behavior of electrons in graphene, 
and their response to various kinds of external perturbations and fields 
\cite{CastroNeto-Guinea-etal-2009}. The Hamiltonian that reflects this physics 
is given by
\begin{equation}
  H = - \sum_{i,\bm{n}} t\bigl(\bX_i,\bX_i+\bm{n}\bigr)
  \, a^\dagger_{\bX_i} b^{\phantom{\dagger}}_{\bX_i+\bm{n}}
  + \textrm{H.~c.}
  \label{ap-eq:Htb}
  .
\end{equation}
The bipartite nature of the honeycomb lattice is evident in this expression by 
the explicit distinction between the lattice sites belonging to sub-lattice $A$ 
or $B$. The second-quantized operator $a_{\bX_i}$($b_{\bX_i}$) destroys an 
electron in a $p_z$ orbital that belongs to a carbon atom located on site
$A$($B$) of the unit cell placed at $\bX_i$. The parameter 
$t\bigl(\bX_i,\bX_i+\bm{n}\bigr)$ is the hopping amplitude between two 
neighboring $\pi$ orbitals, and $\bm{n}$ runs over the three unit cells 
containing a $B$ atom neighboring the $A$ atom from the unit cell at $\bX_i$.
The hopping amplitude is constant in the perfect crystal:
$t\bigl(\bX_i,\bX_i+\bm{n}\bigr)=t_0=2.7$\,eV. But, since $t$ depends strongly 
on the inter-atomic distance, any local change caused by a deformation leads 
to perturbations to this equilibrium value and, hence, more generically, 
$t\bigl(\bX_i,\bX_i+\bm{n}\bigr)=t_0+\delta t\bigl(\bX_i,\bX_i+\bm{n}\bigr)$. 
The presence of $\delta t$, which is a relatively small perturbation to $t_0$ 
in 
practical situations, adds a correction to the low-energy Dirac-like 
Hamiltonian that emerges from (\ref{ap-eq:Htb}) so that the effective 
Hamiltonian 
around the point $\bm{K}=(4\pi/3\sqrt{3}a,\,0)$ in the first Brillouin zone has 
the form \cite{Kane:1997,Suzuura:2002}
\begin{equation}
  \mathcal{H} = v_F\,
  \bm{\sigma}\cdot\bigl(\bm{p}+e\bm{\Acal}\bigr),
  \label{ap-eq:HDirac}
\end{equation}
where $\bm{\sigma}$ is a vector of Pauli matrices, and $\hbar v_F = 3 t_0 a/2$, 
with $a=1.42$\,\AA~the carbon-carbon distance in equilibrium. For deformations 
on scales that are large compared to $a$, the curvature-induced tilting of 
neighboring $p_z$ orbitals can be neglected
\footnote{Note, however, that this is not a restriction on the applicability of 
the method. The assumption of small deviations from the planar configuration is 
for convenience and definiteness only. A full parametrization of the 
hopping modifications including curvature-induced re-hybridization would be 
dealt with in precisely the same way, because the only ingredient that is 
needed is the dependence of the PMF $B$ on the strain components. The central 
and only requirement is the ability to explicitly specify this dependence, as 
done in \Eqref{eq:PMF-def} under the stated conditions.}%
. In this situation the hopping 
amplitude $t$ depends only on the distance between neighboring atoms, and 
we straightforwardly obtain the components of the vector potential 
$\bm{\Acal}=\Acal_{x}\bm{e}_x+\Acal_{y}\bm{e}_y$ by expanding $t$ to 
linear order in the deformation tensor. Choosing the coordinate system so 
that $\bm{e}_x$ is along the zig-zag direction of the honeycomb lattice one 
obtains \cite{Suzuura:2002}
\begin{eqnarray}
  \Acal_x(\bX) - \im \Acal_y(\bX)
  & \equiv -\frac{1}{e v_F} \sum_{\bm{n}} \delta t\bigl(\bX,\bX+\bm{n}\bigr)
  \mathrm{e}^{\im\bm{K}\cdot\bm{n}} \nonumber \\
  & \simeq -\frac{\hbar c}{2 e a}
  \bigl(\varepsilon_{xx} - \varepsilon_{yy} + 2\im\,\varepsilon_{xy}\bigr)
  ,
  \label{ap-eq:Adef}
\end{eqnarray}
where $c = - \partial\log t(r) / \partial\log r\bigr|_{r=a}$. For static 
deformations, a value $c \approx 3.37$ captures the changes in various physical 
properties arising from strain-induced modifications of the $\pi$ bands in 
agreement with first-principles calculations 
\cite{PereiraPRB2009,Ribeiro:2009,Ni:2010,Farjam:2010,Son:2010,
Pereira-OpticalStrain:2010}. Finally, the pseudomagnetic field $B$, being 
defined as the 2D curl of $\bm{\Acal}$, reads: 
\begin{equation}
  B(\bX) = \frac{\hbar c}{ae}
  \sbra{\pd{}{Y}\bra{\frac{\ep_{11}-\ep_{22}}{2}}+\pd{\ep_{12}}{X}}.
  \label{ap-eq:PMF-def}
\end{equation}

As noted in the previous section, by virtue of our choice of piecewise affine 
finite elements for the numerical interpolation, the six variables $v_1$, 
$v_2$, $w$, $M_{11}$, $M_{12}$, and $M_{22}$ are treated as continuous, and 
affine over each triangular element.
As a consequence, from (\ref{ap-strain-vk}) the strain field will be 
discontinuous 
and constant in each triangular element. As the strain components 
$\ep_{\al\be}$ 
are discontinuous under this approximation, the PMF (\ref{ap-eq:PMF-def}) using 
this scheme is undefined.
To overcome this difficulty, we use the technique of patch recovery.
For details of the technique, first described by \citet{Zienkiewicz-Zhu-1992}, 
we refer the reader to \ref{ap:strainrecovery}. In brief, this is a 
mechanism that uses the discontinuous strain data $\ep_{\al\be}$ to recover a 
strain field $\ep_{\al\be}^{\mathrm{rec}}$ of the same type as the primary 
variables: continuous and affine over each element. The derivative of 
$\ep_{\al\be}^{\mathrm{rec}}$ is well-defined, and thus so is the PMF if it is 
calculated using this recovered strain field:
\begin{equation}
  B(\bX) = \frac{\hbar c}{ae} 
  \sbra{\pd{}{Y}\bra{\frac{\ep_{11}^{\mathrm{rec}}-\ep_{22}^{\mathrm{rec}}}{2}} 
  + \pd{\ep_{12}^{\mathrm{rec}}}{X}}
  .
  \label{ap-eq:PMF-def-rec}
\end{equation}

\section{Nondimensionalization}\label{ap:nondimensionalization}

In solving the optimization problem of section \ref{sect:optimization} 
numerically, the first step is to nondimensionalize the 
system of equations in such a way that most variables are $O(1)$ to ensure good 
numerical behavior.
To accomplish this we choose the following scalings, where an overbar 
represents 
the dimensionless quantity.
Set
\begin{equation}
\ep=\frac{aeB_0L}{\hbar c}\label{ap-eq:defn_ep}
\end{equation}
to be the typical scaling of the strain field, then
%
% \numparts
\begin{eqnarray}
(X,Y)=L(\bar{X},\bar{Y}),\quad v_\al=L\ep\bar{v}_\al,\quad 
w=L\sqrt{\ep}\bar{w},\\
\ep_{\al\be}=\ep\bar{\ep}_{\al\be},\quad\ep_{\al\be}^{\mathrm{rec}}=\ep\bar{\ep
}_{\al\be}^{\mathrm{\,rec}},\quad N_{\al\be}=C\ep\bar{N}_{\al\be},\quad 
M_{\al\be}=\frac{D\sqrt{\ep}}{L}\bar{M}_{\al\be},\\
V=C\ep^2\bar{V},\quad B=B_0\bar{B},\quad \hat{B}=B_0\bar{\hat{B}}.
\end{eqnarray}
% \endnumparts
%
All these constants are previously-defined, with the exception 
of $L$, representing the typical size of the domain $\Om$, and $B_0$, the 
typical magnitude of the target PMF $\hat{B}$.
Under these scalings the equations exhibit only one dimensionless parameter, 
namely the dimensionless bending stiffness $\ka$:
\begin{equation}
\ka=\frac{D}{CL^2\ep}=\frac{D\hbar c}{CL^3 aeB_0}.\label{ap-eq:defn_kappa}
\end{equation}

For completeness we will summarize the minimization problem in its 
dimensionless 
form:
\begin{equation}
\textrm{Minimize}\qquad\bar{\mathcal{I}}= 
\int_{\bar{\Om}}\bra{\bar{B}-\bar{\hat{B}}(\bar{X},\bar{Y})}^2\mr{d}^2\bar{\mb{
X}}+\eta\,\bar{\mc{I}}^{\mathrm{reg}}[\la_i],
\end{equation}
subject to the six equations
%
% \numparts
\begin{eqnarray}
\int_{\bar{\Om}}\sbra{\pd{\tilde{v}_1}{\bar{X}}\bar{N}_{11}+\pd{\tilde{v}_1}{
\bar{Y}}\bar{N}_{12}+\tilde{v}_1\bar{V}_{\bar{v}_1}[\bar{w},\bar{v}_1,\bar{v}
_2;\la_i]}\mr{d}^2\bar{\mb{X}}&=0,\label{ap-nondim-weak1}\\
\int_{\bar{\Om}}\sbra{\pd{\tilde{v}_2}{\bar{X}}\bar{N}_{12}+\pd{\tilde{v}_2}{
\bar{Y}}\bar{N}_{22}+\tilde{v}_2\bar{V}_{\bar{v}_2}[\bar{w},\bar{v}_1,\bar{v}
_2;\la_i]}\mr{d}^2\bar{\mb{X}}&=0,\label{ap-nondim-weak2}\\
\int_{\bar{\Om}}
\biggl\{
\pd{\tilde{w}}{\bar{X}}\sbra{-\kappa\bra{\pd{\bar{M}_{11}}{\bar{X}}+\pd{\bar{M}_
{12}}{\bar{Y}}}+\bar{N}_{11}\pd{\bar{w}}{\bar{X}}+\bar{N}_{12}\pd{\bar{w}}{\bar{
Y}}} & \nonumber \\
+\pd{\tilde{w}}{\bar{Y}}\sbra{-\kappa\bra{\pd{\bar{M}_{12}}{\bar{X}}+\pd{
\bar{M}_{22}}{\bar{Y}}}+\bar{N}_{12}\pd{\bar{w}}{\bar{X}}+\bar{N}_{22}\pd{\bar{w
}}{\bar{Y}}} & \nonumber \\
+ \tilde{w}\bar{V}_{\bar{w}}[\bar{w},\bar{v}_1,\bar{v}_2;\la_i]\biggr\} \mr{ 
d } ^2\bar{\mb{X}} & =0,\label{ap-nondim-weak3} \\
\int_{\bar{\Om}}\sbra{\frac{1}{(1-\sigma^2)}(\bar{M}_{11}-\sigma 
\bar{M}_{22})\tilde{M}_{11}+\pd{\bar{w}}{\bar{X}}\pd{\tilde{M}_{11}}{\bar{X}}}
\mr{d}^2\bar{\mb{X}} & =0,\label{ap-nondim-weak4}\\
\int_{\bar{\Om}}\sbra{\frac{1}{(1-\sigma)}\bar{M}_{12}\tilde{M}_{12}+\frac{1}{2
}\pd{\bar{w}}{\bar{X}}\pd{\tilde{M}_{12}}{\bar{Y}}+\frac{1}{2}\pd{\bar{w}}{\bar 
{ 
Y}}\pd{\tilde{M}_{12}}{\bar{X}}}\mr{d}^2\bar{\mb{X}}&=0,\label{ap-nondim-weak5}
\\
\int_{\bar{\Om}}\sbra{\frac{1}{(1-\sigma^2)}(\bar{M}_{22}-\sigma 
\bar{M}_{11})\tilde{M}_{22}+\pd{\bar{w}}{\bar{Y}}\pd{\tilde{M}_{22}}{\bar{Y}}}
\mr{d}^2\bar{\mb{X}}&=0,\label{ap-nondim-weak6}
\end{eqnarray}
% \endnumparts
%
together with the additional definitions
%
% \numparts
\begin{eqnarray}
\bar{N}_{11}=\bar{\ep}_{11}+\nu\bar{\ep}_{22},\qquad 
\bar{N}_{12}=(1-\nu)\bar{\ep}_{12},\qquad 
\bar{N}_{22}=\nu\bar{\ep}_{11}+\bar{\ep}_{22},\\
\bar{\ep}_{11}=\pd{\bar{v}_1}{\bar{X}}+\frac{1}{2}\bra{\pd{\bar{w}}{\bar{X}}}^2 
\!\!,\quad 
\bar{\ep}_{22}=\pd{\bar{v}_2}{\bar{Y}}+\frac{1}{2}\bra{\pd{\bar{w}}{\bar{Y}}}^2,
\\
\bar{\ep}_{12}=\frac{1}{2}\bra{\pd{\bar{v}_1}{\bar{Y}}+\pd{\bar{v}_2}{\bar{X}}
+\pd{\bar{w}}{\bar{X}}\pd{\bar{w}}{\bar{Y}}},\\
\bar{B}=\pd{}{\bar{Y}}\bra{\frac{\bar{\ep}_{11}^{\mathrm{\,rec}}-\bar{\ep}_{22}^
{ \mathrm{\,rec}}}{2}}+\pd{\bar{\ep}_{12}^{\mathrm{\,rec}}}{\bar{X}},
\end{eqnarray}
% \endnumparts
%
and $\bar{\ep}^{\mathrm{rec}}_{\al\be}$ obtained from $\bar{\ep}_{\al\be}$ 
by strain recovery.

To make the expression (\ref{eq:interaction-potential}) for the 
substrate--graphene interaction dimensionless, set
\begin{equation}
\fl (x,y)=L(\bar{x},\bar{y}),\quad s=L\sqrt{\ep}\bar{s},\quad 
\hat{z}(x,y)=L\sqrt{\ep}\bar{\hat{z}}(\bar{x},\bar{y}),\quad 
J(s)=J_0\bar{J}(\bar{s}),
\end{equation}
so that
\begin{eqnarray}
\bar{V}[\bar{w},\bar{v}_1,\bar{v}_2;\la_i]=\bar{p}\bar{w}+\bar{J}_0\bar{J}
\bigl(\bar { w}-\bar{\hat{z}}(\bar{X}+\ep\bar{v}_1,\bar{Y}+\ep\bar{v}_2)\bigr),
\end{eqnarray}
where
\begin{eqnarray}
\bar{J}(\bar{s})=\frac{1}{3}\sbra{\bra{\frac{\bar{s}^*}{\bar{s}}}^8-4\bra{\frac{
\bar{s}^*}{\bar{s}}}^2},
\end{eqnarray}
and the three dimensionless parameters are given by
\begin{eqnarray}
\bar{p}=\frac{pL}{C\ep^{3/2}},\qquad \bar{J}_0=\frac{J_0}{C\ep^2},\qquad 
\bar{s}^*=\frac{s^*}{L\sqrt{\ep}}.
\end{eqnarray}
As a representative value for $s^*$, we use the value that 
\citet{Xu-Buehler-2010} give for C--Cu, namely $2.243\,\textrm{\AA}$.
Similarly we use $J_0=0.45\,\mathrm{J\,m^{-2}}$ as a representative value, from 
the investigation of \citet{Koenig-Boddeti-etal-2011} into the adhesion 
strength 
between graphene and SiO${}_2$. We select a typical value for the hydrostatic 
pressure as $p=100\,\textrm{bar}=10^7\,\mathrm{Pa}$.
These values, for a lengthscale $L=10^{-8}\,\mathrm{m}$ and a target PMF scale 
$B_0=10\,\mathrm{T}$, give $\bar{p}=0.545$, $\bar{J}_0=30.66$ and 
$\bar{s}^*=0.280$.

The derivatives of the potentials appearing in the dimensionless weak form 
equations are
\begin{eqnarray}
\fl \bar{V}_{\bar{w}}[\bar{w},\bar{v}_1,\bar{v}_2;\la_i]=\bar{p}+\bar{J}_0 
\bar{J}'(\bar{w}-\bar{\hat{z}}(\bar{X}+\ep \bar{v}_1,\bar{Y}+\ep 
\bar{v}_2)),\label{Vw}\\
\fl \bar{V}_{\bar{v}_\al}[\bar{w},\bar{v}_1,\bar{v}_2;\la_i]=-\ep \bar{J}_0 
\bar{J}'(\bar{w}-\bar{\hat{z}}(\bar{X}+\ep \bar{v}_1,\bar{Y}+\ep 
\bar{v}_2))\at{\pd{\bar{\hat{z}}}{\bar{x}_\al}}{(\bar{X}+\ep 
\bar{v}_1,\bar{Y}+\ep \bar{v}_2)},\label{Vva}
\end{eqnarray}
where
\begin{eqnarray}
\bar{J}'(\bar{s})=\frac{8}{3\bar{s}^*}\sbra{\bra{\frac{\bar{s}^*}{\bar{s}}}
^3-\bra{\frac{\bar{s}^*}{\bar{s}}}^9}.\label{Jp-def}
\end{eqnarray}
In section \ref{sect:substrate-parametrization} and subsequently, all variables 
are assumed to be dimensionless, and \emph{overbars 
are omitted for clarity}.

\section{Parametrization of the substrate topography}
\label{ap:substrate-topography}

% ------------------------------------------------------------------------------
% FIGURE
% ------------------------------------------------------------------------------
\begin{figure}[tb]
  \centering
  \includegraphics[width=0.3\textwidth]{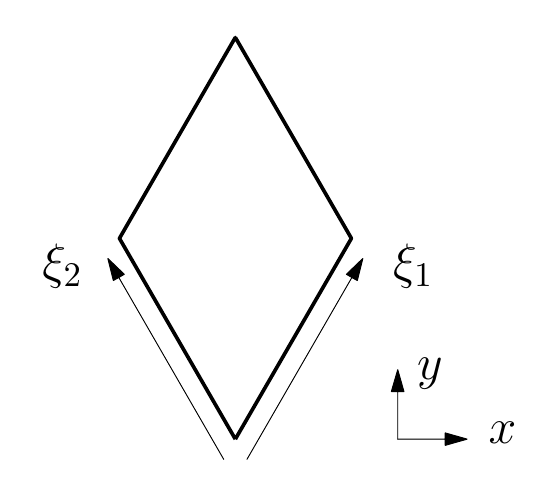}%
  \caption{
    The repeating unit cell, defined using the variables $\xi_1$ and $\xi_2$.
  }
  \label{fig:unitrhombus}
\end{figure}
% ------------------------------------------------------------------------------

For the numerical experiments in this article, we assume that the substrate is 
patterned periodically in the two horizontal directions, and set the repeating 
unit cell to be the rhombus depicted in Figure~\ref{fig:unitrhombus}. If we 
introduce the two coordinates
\begin{eqnarray}
\xi_1=\frac{y}{\sqrt{3}}+x,\qquad \xi_2=\frac{y}{\sqrt{3}}-x,
\end{eqnarray}
the unit cell corresponds to the domain
\begin{eqnarray}
\Om=\left\{(\xi_1,\xi_2)\;:\;0\leq\xi_1<1,\;0\leq\xi_2<1\right\}.
\end{eqnarray}
The topography of the substrate, $\hat{z}(x,y)$, can then be resolved in terms 
of a sum of functions which are periodic on the unit cell $\Om$. Such functions 
take one of the four following forms:
%
% \numparts
\begin{eqnarray}
f_{kl}^1=\cos(2\pi k\xi_1)\,\cos(2\pi l\xi_2),\\
f_{kl}^2=\cos(2\pi k\xi_1)\,\sin(2\pi l\xi_2),\\
f_{kl}^3=\sin(2\pi k\xi_1)\,\cos(2\pi l\xi_2),\\
f_{kl}^4=\sin(2\pi k\xi_1)\,\sin(2\pi l\xi_2),
\end{eqnarray}
% \endnumparts
%
and so we choose a truncated expansion in terms of these as follows:
\begin{eqnarray}
\hat{z}(x,y)=\sum_{k=0}^K\sum_{l=0}^K\bra{\al_{kl}f_{kl}^1+ 
\be_{kl}f_{kl}^2+\ga_{kl}f_{kl}^3+\de_{kl}f_{kl}^4},
\end{eqnarray}
with the constants $\al_{kl}$, $\be_{kl}$, $\ga_{kl}$, $\de_{kl}$ playing the 
role of the control variables $\la_i$: varying the topography of the 
substrate is, therefore, achieved by varying these $4(K+1)^2$ constants.
For convenience we set
\begin{eqnarray}
\be_{k0}=\ga_{0k}=\de_{k0}=\de_{0k}=0
\end{eqnarray}
for each $k$ and, since rigid vertical displacements of the substrate do not 
affect the objective function, we further set $\al_{00}=0$.

In order to avoid convergence towards solutions that are 
ill-behaved (in this case those could be, for example, substrate profiles with 
discontinuities or sharp topographical features) during the numerical 
optimization, a regularization term, $\mc{I}^{\mathrm{reg}}$, is added 
to the objective integral, as per \Eqref{eq:objfun}. We choose it to be
\begin{eqnarray}
\mathcal{I}^{\mathrm{reg}}&=\frac{1}{\textrm{Area}(\Om)}\int_\Om|\mb{\nabla}
\hat{z}|^2\mathrm{d}^2\mb{X} \nonumber\\
&=\frac{4}{3}\int_0^1\int_0^1\sbra{\bra{\pd{\hat{z}}{\xi_1}}^2 
+\bra{\pd{\hat{z}}{\xi_2}}^2 - \pd{\hat{z}}{ 
\xi_1}\pd{\hat{z}}{\xi_2}}\,\mr{d}\xi_1\mr{d}\xi_2,
\end{eqnarray}
which is simple to calculate using the orthogonality of the basis functions 
$f_{kl}^n$ over $\Om$.

The domain $\Om$ for the graphene sheet will also be the unit rhombus, with 
periodic boundary conditions applied to all six state variables.
However, we set the displacement components $v_1=v_2=0$ at the corner points to 
disallow arbitrarily-large horizontal rigid displacements. This is a reasonable 
constraint on account of the two-dimensional periodicity of the substrate.
Due to the geometry of the unit rhombus, we can set the triangulation to be a 
regular isometric grid.

\section{Strain recovery}\label{ap:strainrecovery}

\begin{figure}[tb]
  \centering
  \includegraphics[width=0.85\columnwidth]{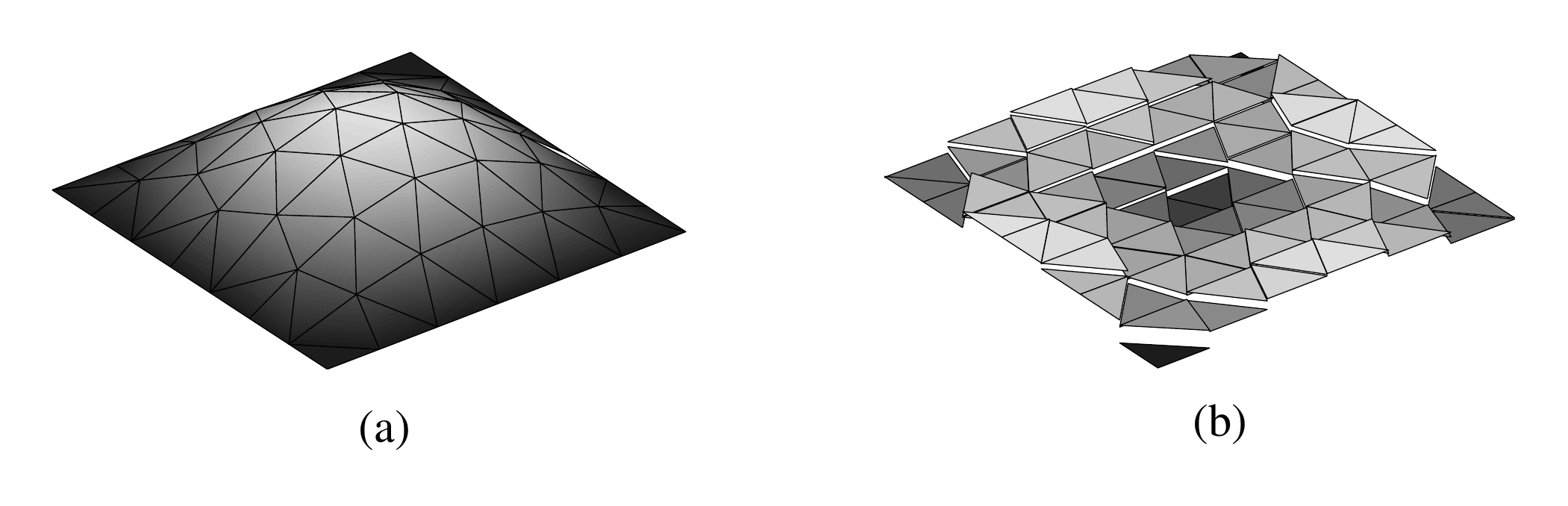}%
  \caption{
    Surface plots of (a) a continuous, piecewise affine function $f$ defined on 
    a triangulated domain, and (b) its gradient $|\mb{\nabla}f|$, which is 
    discontinuous and constant over each triangular element.
  }
  \label{fig:tri_finite_element}
\end{figure}

As noted in \ref{ap:elasticity}, we choose a finite element 
discretization for our six variables $v_1$, $v_2$, $w$, $M_{11}$, $M_{12}$, 
$M_{22}$ that approximates these quantities with functions that are continuous 
across the domain $\Om$, and affine over each triangular element in the 
discretization (see Figure~\ref{fig:tri_finite_element}(a) for a representation 
of such a function). 
Thus the quantities can be parametrized by their values at each nodal point of 
the triangulation. 
Differentiating such a function leads to a discontinuous function, which is 
constant on each triangular element, as shown in Figure~\ref{fig:tri_finite_element}(b). 
 
% ------------------------------------------------------------------------------
% FIGURE
% ------------------------------------------------------------------------------
% ------------------------------------------------------------------------------

The piecewise constant function is a less accurate approximation than the 
continuous piecewise affine function, and this led to the patch recovery method 
\citep{Zienkiewicz-Zhu-1992}, which reconstructs an accurate continuous 
piecewise affine representation of a quantity calculated as a piecewise constant 
function. 
The canonical example where this recovery method becomes relevant is in 
elasticity with piecewise affine displacements leading to a piecewise constant 
stress field.
The original purpose of the patch recovery method was to find a better 
approximation to the stress field calculated from a displacement-based finite 
element method.
In this article a piecewise affine displacement field leads to a piecewise 
constant approximation $\ep_{\al\be}$ to the strain field, whereas we require a 
differentiable approximation. By using the patch recovery method we recover a 
piecewise affine strain field, 
$\ep_{\al\be}^{\mathrm{rec}}$, which we are able to differentiate to find the 
PMF $B$ according to the prescription in \Eqref{ap-eq:PMF-def-rec}.

To illustrate the patch recovery method, consider a triangulation of the domain 
$\Om$ which defines the spatial extent of the medium with triangles 
$k=1,\ldots,N_t$ and nodes $i=1,\ldots,N_p$. We have a function $f$, constant on 
each element (so $f(X,Y)=f_k$ if $\mb{X}$ is in triangle $k$), from which we 
want to recover a piecewise affine function $f^{\mathrm{rec}}$ (defined by its 
values $f_i^{\mathrm{rec}}$ at each node $\mb{X}_i$ of the triangulation).

% ------------------------------------------------------------------------------
% FIGURE
% ------------------------------------------------------------------------------
\begin{figure}[tb]
  \centering
  \includegraphics[width=0.4\textwidth]{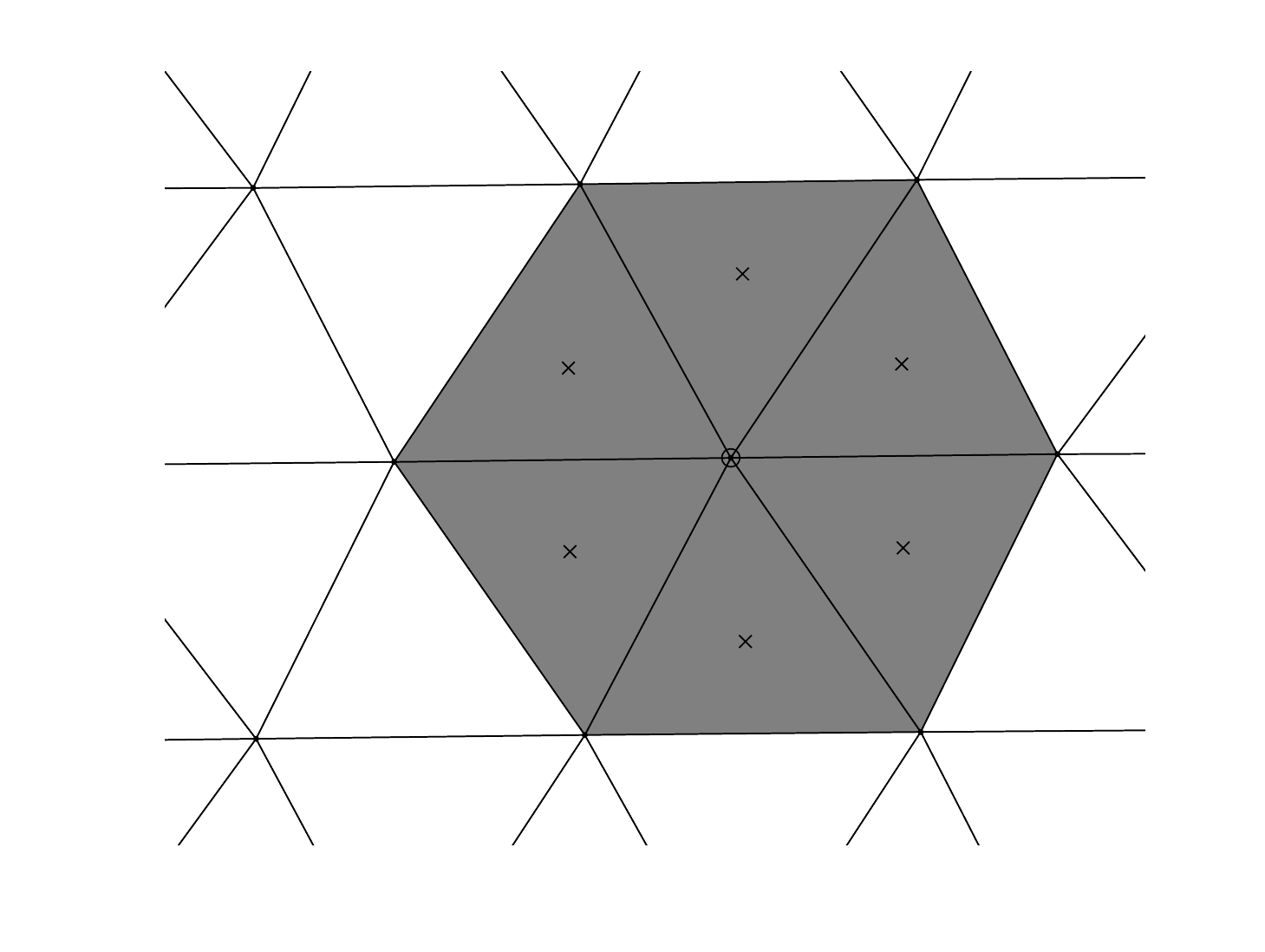}
  \caption{The patch of elements surrounding a node ($\circ$) in a 
    triangulation. An affine function $f^{\mathrm{fit}}$ is fit to the patch 
    using the values of the original function $f$ evaluated at the element 
    centroids $\times$.
  }
  \label{fig:patch}
\end{figure}
% ------------------------------------------------------------------------------

The strength of the patch recovery method is that the nodal values 
$f_i^{\mathrm{rec}}$ are calculated individually in turn, rather than in a 
global optimization over all values at once. 
For each nodal point $i$, we identify the patch, which (for triangular elements 
with a piecewise affine target) is the set of all elements that contain the node 
$i$ as a vertex, as displayed in Figure~\ref{fig:patch}. 
The key step in the process is to fit a function $f_i^{\mathrm{fit}}(X,Y)$ to 
the patch for node $i$ that is of the same order as the proposed target 
function.
So, in this case, we need to fit an affine function 
$f_i^{\mathrm{fit}}(X,Y)=a+bX+cY$ to the patch.
We use the values $f_k$, evaluated at the centroids $(X_k^c,Y_k^c)$ of the 
elements, to calculate the parameters $(a,b,c)=\mb{a}^T$ through a least-squares 
optimization. 
As noted in \citet{Zienkiewicz-Zhu-1992}, $\mb{a}$ is thus found by solving the 
system
\begin{eqnarray}
\sbra{\sum_{k\in\mathrm{patch}(i)}\mb{p}_k\,\mb{p}_k^T}\mb{a}=\sum_{k\in\mathrm{
patch}(i)}f_k \, \mb{p}_k,\label{ap-eq:patchrecovery}
\end{eqnarray}
where $\mb{p}_k=(1,X_k^c,Y_k^c)^T$.
Having found $f_i^{\mathrm{fit}}(X,Y)$, the nodal value of the recovered 
function $f^{\mathrm{rec}}$ is simply 
$f_i^{\mathrm{rec}}=f_i^{\mathrm{fit}}(X_i,Y_i)$, the fit function evaluated at 
the nodal point. 
At the domain boundaries there will usually be too few elements in the patch for 
the system (\ref{ap-eq:patchrecovery}) to be well-conditioned.
Instead, we would follow \citet{Zienkiewicz-Zhu-1992} and find the boundary 
nodal values of $f^{\mathrm{rec}}$ by using the interior patches, and average 
over all the calculated values.
This consideration does not apply for periodic boundary conditions, since in 
that case we can treat the entire domain as being of infinite extent, and all 
points are interior points.

\section{Three-dimensional plots}\label{app:3dplots}
\begin{figure}[!p]
\centering
\includegraphics[height=0.21\textheight]{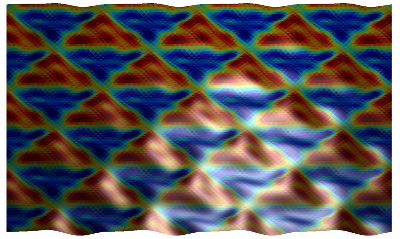}\\[1em]
\includegraphics[height=0.21\textheight]{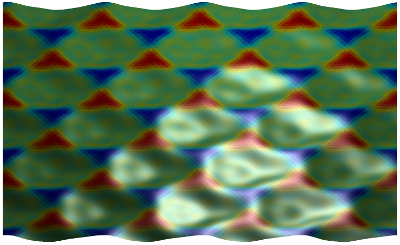}\\[1em]
\includegraphics[height=0.21\textheight]{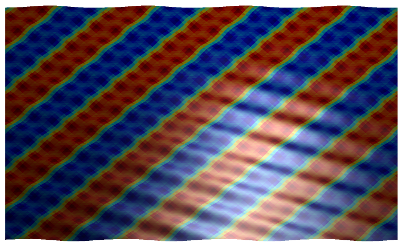}\\[1em]
\includegraphics[height=0.21\textheight]{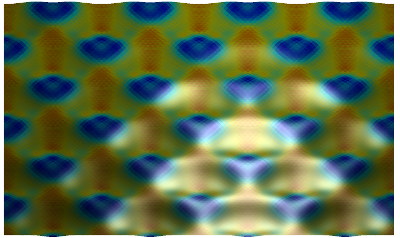}
\caption{Three-dimensional plots of graphene sheets deformed by the four substrates in Figure \ref{fig:B10_results}, colored with the resultant pseudomagnetic fields. Vertical scales in some of the plots are exaggerated for clarity, by factors of 2, 3, 1 and 3 respectively.}
\label{SIfig:3dplot-B10}
\end{figure}

Figure \ref{SIfig:3dplot-B10} shows three-dimensional visualizations of deformed graphene sheets corresponding to the four solutions of Figure \ref{fig:B10_results}, where the target PMF value was $10\,\mathrm{T}$. Vertical scales are exaggerated for clarity.

\begin{figure}[!p]
\centering
\includegraphics[height=0.21\textheight]{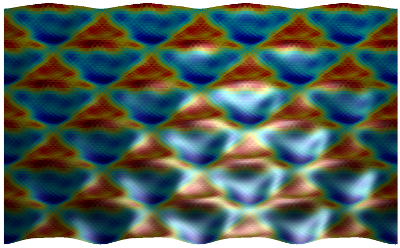}\\[1em]
\includegraphics[height=0.21\textheight]{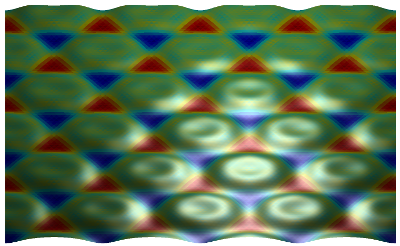}\\[1em]
\includegraphics[height=0.21\textheight]{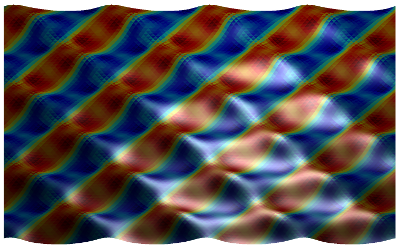}\\[1em]
\includegraphics[height=0.21\textheight]{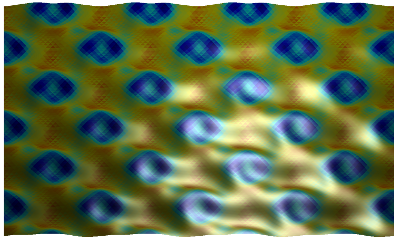}
\caption{Three-dimensional plots of graphene sheets deformed by the four substrates in Figure \ref{fig:B100_results}, colored with the resultant pseudomagnetic fields. Vertical scales are not exaggerated.}
\label{SIfig:3dplot-B100}
\end{figure}

The corresponding visualizations for $B=100\,\mathrm{T}$ (corresponding to Figure \ref{fig:B100_results}) are shown in Figure \ref{SIfig:3dplot-B100}. In this case the vertical scale is not exaggerated.

\clearpage

% ------------------------------------------------------------------------------
% SUPPLEMENTARY REFERENCES
% ------------------------------------------------------------------------------
% \clearpage
% \def\bibsection{\section*{Supplementary \refname}}
% \putbib[graphrefs]
% \end{bibunit}

\def\newblock{\hskip .11em plus .33em minus .07em}
\bibliography{graphrefs}

\begin{thebibliography}{75}
\providecommand{\natexlab}[1]{#1}
\providecommand{\url}[1]{\texttt{#1}}
\expandafter\ifx\csname urlstyle\endcsname\relax
  \providecommand{\doi}[1]{doi: #1}\else
  \providecommand{\doi}{doi: \begingroup \urlstyle{rm}\Url}\fi

\bibitem[Lee et~al.(2008)Lee, Wei, Kysar, and Hone]{LeeSCIENCE2008}
C.~Lee, X.~Wei, J.W. Kysar, and J.~Hone.
\newblock {Measurement of the elastic properties and intrinsic strength of
  monolayer graphene}.
\newblock \emph{Science}, 321\penalty0 (5887):\penalty0 385--388, 2008.

\bibitem[Klimov et~al.(2012)Klimov, Jung, Zhu, Li, Wright, Solares, Newell,
  Zhitenev, and Stroscio]{Klimov-Jung-etal-2012}
N.N. Klimov, S.~Jung, S.~Zhu, T.~Li, C.A. Wright, S.D. Solares, D.B. Newell,
  N.B. Zhitenev, and J.A. Stroscio.
\newblock {Electromechanical properties of graphene drumheads}.
\newblock \emph{Science}, 336\penalty0 (6088):\penalty0 1557--1561, 2012.

\bibitem[Xu et~al.(2012)Xu, Yang, Barber, Ackerman, Schoelz, Qi, Kornev, Dong,
  Bellaiche, Barraza-Lopez, and Thibado]{Xu-Yang-etal-2012}
P.~Xu, Y.~Yang, S.D. Barber, M.L. Ackerman, J.K. Schoelz, D.~Qi, I.A. Kornev,
  L.~Dong, L.~Bellaiche, S.~Barraza-Lopez, and P.M. Thibado.
\newblock {Atomic control of strain in freestanding graphene}.
\newblock \emph{Phys. Rev. B}, 85\penalty0 (12):\penalty0 121406, 2012.

\bibitem[Metzger et~al.(2010)Metzger, R\'{e}mi, Liu, Kusminskiy, {Castro Neto},
  Swan, and Goldberg]{Metzger2010Biaxial}
C.~Metzger, S.~R\'{e}mi, M.~Liu, S.V. Kusminskiy, A.H. {Castro Neto}, A.K.
  Swan, and B.B. Goldberg.
\newblock {Biaxial strain in graphene adhered to shallow depressions}.
\newblock \emph{Nano Lett.}, 10\penalty0 (1):\penalty0 6--10, 2010.

\bibitem[Scharfenberg et~al.(2011)Scharfenberg, Rocklin, Chialvo, Weaver,
  Goldbart, and Mason]{Scharfenberg2011Probing}
S.~Scharfenberg, D.Z. Rocklin, C.~Chialvo, R.L. Weaver, P.M. Goldbart, and
  N.~Mason.
\newblock {Probing the mechanical properties of graphene using a corrugated
  elastic substrate}.
\newblock \emph{Appl. Phys. Lett.}, 98\penalty0 (9):\penalty0 091908, 2011.

\bibitem[Tomori et~al.(2011)Tomori, Kanda, Goto, Ootuka, Tsukagoshi, Moriyama,
  Watanabe, and Tsuya]{Tomori-Kanda-etal-2011}
H.~Tomori, A.~Kanda, H.~Goto, Y.~Ootuka, K.~Tsukagoshi, S.~Moriyama,
  E.~Watanabe, and D.~Tsuya.
\newblock {Introducing nonuniform strain to graphene using dielectric
  nanopillars}.
\newblock \emph{Appl. Phys. Express}, 4\penalty0 (7):\penalty0 075102, 2011.

\bibitem[Koenig et~al.(2011)Koenig, Boddeti, Dunn, and
  Bunch]{Koenig-Boddeti-etal-2011}
S.P. Koenig, N.G. Boddeti, M.L. Dunn, and J.S. Bunch.
\newblock {Ultrastrong adhesion of graphene membranes}.
\newblock \emph{Nat. Nanotechnol.}, 6\penalty0 (9):\penalty0 543--546, 2011.

\bibitem[Bao et~al.(2009)Bao, Miao, Chen, Zhang, Jang, Dames, and
  Lau]{Bao-Miao-etal-2009}
W.~Bao, F.~Miao, Z.~Chen, H.~Zhang, W.~Jang, C.~Dames, and C.N. Lau.
\newblock {Controlled ripple texturing of suspended graphene and ultrathin
  graphite membranes}.
\newblock \emph{Nat. Nanotechnol.}, 4\penalty0 (9):\penalty0 562--566, 2009.

\bibitem[Georgiou et~al.(2011)Georgiou, Britnell, Blake, Gorbachev, Gholinia,
  Geim, Casiraghi, and Novoselov]{Georgiou-Britnell-etal-2011}
T.~Georgiou, L.~Britnell, P.~Blake, R.V. Gorbachev, A.~Gholinia, A.K. Geim,
  C.~Casiraghi, and K.S. Novoselov.
\newblock {Graphene bubbles with controllable curvature}.
\newblock \emph{Appl. Phys. Lett.}, 99\penalty0 (9):\penalty0 093103, 2011.

\bibitem[Zang et~al.(2013)Zang, Ryu, Pugno, Wang, Tu, Buehler, and
  Zhao]{Zang2013Multifunctionality}
J.~Zang, S.~Ryu, N.~Pugno, Q.~Wang, Q.~Tu, M.J. Buehler, and X.~Zhao.
\newblock {Multifunctionality and control of the crumpling and unfolding of
  large-area graphene}.
\newblock \emph{Nat. Mater.}, 12\penalty0 (4):\penalty0 321--325, 2013.

\bibitem[Schmidt(2012)]{Schmidt2012Bioelectronics}
C.~Schmidt.
\newblock {Bioelectronics: the bionic material.}
\newblock \emph{Nature}, 483\penalty0 (7389):\penalty0 S37, 2012.

\bibitem[Mannoor et~al.(2012)Mannoor, Tao, Clayton, Sengupta, Kaplan, Naik,
  Verma, Omenetto, and McAlpine]{Mannoor2012Graphenebased}
M.S. Mannoor, H.~Tao, J.D. Clayton, A.~Sengupta, D.L. Kaplan, R.R. Naik,
  N.~Verma, F.G. Omenetto, and M.C. McAlpine.
\newblock {Graphene-based wireless bacteria detection on tooth enamel}.
\newblock \emph{Nat. Commun.}, 3:\penalty0 763, 2012.

\bibitem[Rasool et~al.(2013)Rasool, Ophus, Klug, Zettl, and
  Gimzewski]{Rasool2013Measurement}
H.I. Rasool, C.~Ophus, W.S. Klug, A.~Zettl, and J.K. Gimzewski.
\newblock {Measurement of the intrinsic strength of crystalline and
  polycrystalline graphene}.
\newblock \emph{Nat. Commun.}, 4:\penalty0 2811, 2013.

\bibitem[Zhang et~al.(2014{\natexlab{a}})Zhang, Ma, Fan, Zeng, Peng, Loya, Liu,
  Gong, Zhang, Zhang, Ajayan, Zhu, and Lou]{Zhang2014Fracture}
P.~Zhang, L.~Ma, F.~Fan, Z.~Zeng, C.~Peng, P.E. Loya, Z.~Liu, Y.~Gong,
  J.~Zhang, X.~Zhang, P.M. Ajayan, T.~Zhu, and J.~Lou.
\newblock {Fracture toughness of graphene}.
\newblock \emph{Nat. Commun.}, 5:\penalty0 3782, 2014{\natexlab{a}}.

\bibitem[Kane and Mele(1997)]{Kane:1997}
C.L. Kane and E.J. Mele.
\newblock {Size, shape, and low energy electronic structure of carbon
  nanotubes}.
\newblock \emph{Phys. Rev. Lett.}, 78\penalty0 (10):\penalty0 1932--1935, 1997.

\bibitem[Suzuura and Ando(2002)]{Suzuura:2002}
H.~Suzuura and T.~Ando.
\newblock {Phonons and electron-phonon scattering in carbon nanotubes}.
\newblock \emph{Phys. Rev. B}, 65\penalty0 (23):\penalty0 235412, 2002.

\bibitem[{Castro Neto} et~al.(2009){Castro Neto}, Guinea, Peres, Novoselov, and
  Geim]{CastroNeto-Guinea-etal-2009}
A.H. {Castro Neto}, F.~Guinea, N.M.R. Peres, K.S. Novoselov, and A.K. Geim.
\newblock {The electronic properties of graphene}.
\newblock \emph{Rev. Mod. Phys.}, 81\penalty0 (1):\penalty0 109--162, 2009.

\bibitem[Vozmediano et~al.(2010)Vozmediano, Katsnelson, and
  Guinea]{Vozmediano-Katsnelson-Guinea-2010}
M.A.H. Vozmediano, M.I. Katsnelson, and F.~Guinea.
\newblock {Gauge fields in graphene}.
\newblock \emph{Phys. Rep.}, 496\penalty0 (4-5):\penalty0 109--148, 2010.

\bibitem[Levy et~al.(2010)Levy, Burke, Meaker, Panlasigui, Zettl, Guinea,
  {Castro Neto}, and Crommie]{Levy-Burke-etal-2010}
N.~Levy, S.A. Burke, K.L. Meaker, M.~Panlasigui, A.~Zettl, F.~Guinea, A.H.
  {Castro Neto}, and M.F. Crommie.
\newblock {Strain-induced pseudo-magnetic fields greater than 300 Tesla in
  graphene nanobubbles}.
\newblock \emph{Science}, 329\penalty0 (5991):\penalty0 544--547, 2010.

\bibitem[Lu et~al.(2012)Lu, {Castro Neto}, and Loh]{LuNatureComm2012}
J.~Lu, A.H. {Castro Neto}, and K.P. Loh.
\newblock {Transforming Moir\'{e} blisters into geometric graphene
  nano-bubbles}.
\newblock \emph{Nat. Commun.}, 3\penalty0 (5):\penalty0 823, 2012.

\bibitem[Jang et~al.(2014)Jang, Kim, Shin, Wang, Jang, Kim, Lee, Kim, Song, and
  Kahng]{Jang2014Observation}
W.-J. Jang, H.~Kim, Y.-R. Shin, M.~Wang, S.K. Jang, M.~Kim, S.~Lee, S.-W. Kim,
  Y.J. Song, and S.-J. Kahng.
\newblock {Observation of spatially-varying Fermi velocity in strained-graphene
  directly grown on hexagonal boron nitride}.
\newblock \emph{Carbon}, 74:\penalty0 139--145, 2014.

\bibitem[de~Juan et~al.(2012)de~Juan, Sturla, and Vozmediano]{deJuan2012Space}
F.~de~Juan, M.~Sturla, and M.A.H. Vozmediano.
\newblock {Space dependent Fermi velocity in strained graphene}.
\newblock \emph{Phys. Rev. Lett.}, 108\penalty0 (22):\penalty0 227505, 2012.

\bibitem[de~Juan et~al.(2013)de~Juan, Ma\~{n}es, and
  Vozmediano]{deJuan2013Gauge}
F.~de~Juan, J.L. Ma\~{n}es, and M.A.H. Vozmediano.
\newblock {Gauge fields from strain in graphene}.
\newblock \emph{Phys. Rev. B}, 87\penalty0 (16):\penalty0 165131, 2013.

\bibitem[{Pacheco Sanjuan} et~al.(2014){Pacheco Sanjuan}, Wang, Imani,
  Vanevi\'{c}, and Barraza-Lopez]{BarrazaLopez-Discrete}
A.A. {Pacheco Sanjuan}, Z.~Wang, H.P. Imani, M.~Vanevi\'{c}, and
  S.~Barraza-Lopez.
\newblock {Graphene's morphology and electronic properties from discrete
  differential geometry}.
\newblock \emph{Phys. Rev. B}, 89\penalty0 (12):\penalty0 121403, 2014.

\bibitem[Pereira and {Castro Neto}(2009)]{Pereira-CastroNeto-2009}
V.M. Pereira and A.H. {Castro Neto}.
\newblock {Strain engineering of graphene's electronic structure}.
\newblock \emph{Phys. Rev. Lett.}, 103\penalty0 (4):\penalty0 046801, 2009.

\bibitem[Guinea et~al.(2009)Guinea, Katsnelson, and
  Geim]{Guinea-Katsnelson-Geim-2009}
F.~Guinea, M.I. Katsnelson, and A.K. Geim.
\newblock {Energy gaps and a zero-field quantum Hall effect in graphene by
  strain engineering}.
\newblock \emph{Nat. Phys.}, 6\penalty0 (1):\penalty0 30--33, 2009.

\bibitem[Lim et~al.(2012)Lim, Koon, Zhan, Shen, \"{O}zyilmaz, and
  Sow]{Lim-Koon-etal-2012}
S.X. Lim, G.K.W. Koon, D.~Zhan, Z.~Shen, B.~\"{O}zyilmaz, and C.~Sow.
\newblock {Assembly of suspended graphene on carbon nanotube scaffolds with
  improved functionalities}.
\newblock \emph{Nano Res.}, 5\penalty0 (11):\penalty0 783--795, 2012.

\bibitem[Guinea(2012)]{Guinea-2012}
F.~Guinea.
\newblock {Strain engineering in graphene}.
\newblock \emph{Solid State Commun.}, 152\penalty0 (15):\penalty0 1437--1441,
  2012.

\bibitem[Gomes et~al.(2012)Gomes, Mar, Ko, Guinea, and
  Manoharan]{Gomes2012Designer}
K.K. Gomes, W.~Mar, W.~Ko, F.~Guinea, and H.C. Manoharan.
\newblock {Designer Dirac fermions and topological phases in molecular
  graphene}.
\newblock \emph{Nature}, 483\penalty0 (7389):\penalty0 306--310, 2012.

\bibitem[Guinea et~al.(2010)Guinea, Geim, Katsnelson, and
  Novoselov]{Guinea-Geim-etal-2010}
F.~Guinea, A.K. Geim, M.I. Katsnelson, and K.S. Novoselov.
\newblock {Generating quantizing pseudomagnetic fields by bending graphene
  ribbons}.
\newblock \emph{Phys. Rev. B}, 81\penalty0 (3):\penalty0 035408, 2010.

\bibitem[Low and Guinea(2010)]{low_strain-induced_2010}
T.~Low and F.~Guinea.
\newblock {Strain-induced pseudomagnetic field for novel graphene electronics}.
\newblock \emph{Nano Lett.}, 10\penalty0 (9):\penalty0 3551--3554, 2010.

\bibitem[Kim et~al.(2011)Kim, Blanter, and Ahn]{Kim-Blanter-Ahn-2011}
K.-J. Kim, Y.M. Blanter, and K.-H. Ahn.
\newblock {Interplay between real and pseudomagnetic field in graphene with
  strain}.
\newblock \emph{Phys. Rev. B}, 84\penalty0 (8):\penalty0 081401, 2011.

\bibitem[Neek-Amal et~al.(2012)Neek-Amal, Covaci, and
  Peeters]{NeekAmal-Covaci-Peeters-2012}
M.~Neek-Amal, L.~Covaci, and F.M. Peeters.
\newblock {Nanoengineered nonuniform strain in graphene using nanopillars}.
\newblock \emph{Phys. Rev. B}, 86\penalty0 (4):\penalty0 041405, 2012.

\bibitem[Neek-Amal and Peeters(2012)]{NeekAmal-Peeters-2012}
M.~Neek-Amal and F.M. Peeters.
\newblock {Strain-engineered graphene through a nanostructured substrate. II.
  Pseudomagnetic fields}.
\newblock \emph{Phys. Rev. B}, 85\penalty0 (19):\penalty0 195446, 2012.

\bibitem[Sloan et~al.(2013)Sloan, Sanjuan, Wang, Horvath, and
  Barraza-Lopez]{Barraza-Lopez:2013}
J.V. Sloan, A.A.P. Sanjuan, Z.~Wang, C.~Horvath, and S.~Barraza-Lopez.
\newblock {Strain gauge fields for rippled graphene membranes under central
  mechanical load: an approach beyond first-order continuum elasticity}.
\newblock \emph{Phys. Rev. B}, 87\penalty0 (15):\penalty0 155436, 2013.

\bibitem[Li et~al.(2012)Li, Chung, Chen, and Cheng]{Li2012Nanoscale}
J.~Li, T.-F. Chung, Y.P. Chen, and G.J. Cheng.
\newblock {Nanoscale strainability of graphene by laser shock-induced
  three-dimensional shaping}.
\newblock \emph{Nano Lett.}, 12\penalty0 (9):\penalty0 4577--4583, 2012.

\bibitem[Shioya et~al.(2014)Shioya, Craciun, Russo, Yamamoto, and
  Tarucha]{Shioya2014Straining}
H.~Shioya, M.F. Craciun, S.~Russo, M.~Yamamoto, and S.~Tarucha.
\newblock {Straining graphene using thin film shrinkage methods}.
\newblock \emph{Nano Lett.}, 14\penalty0 (3):\penalty0 1158--1163, 2014.

\bibitem[Wei et~al.(2009)Wei, Fragneaud, Marianetti, and
  Kysar]{Wei-Fragneaud-etal-2009}
X.~Wei, B.~Fragneaud, C.A. Marianetti, and J.W. Kysar.
\newblock {Nonlinear elastic behavior of graphene: ab initio calculations to
  continuum description}.
\newblock \emph{Phys. Rev. B}, 80\penalty0 (20):\penalty0 205407, 2009.

\bibitem[Blakslee et~al.(1970)Blakslee, Proctor, Seldin, Spence, and
  Weng]{Blakslee:1970}
O.L. Blakslee, D.G. Proctor, E.J. Seldin, G.B. Spence, and T.~Weng.
\newblock {Elastic constants of compression-annealed pyrolytic graphite}.
\newblock \emph{J. Appl. Phys.}, 41\penalty0 (8):\penalty0 3373, 1970.

\bibitem[Kudin et~al.(2001)Kudin, Scuseria, and
  Yakobson]{Kudin-Scuseria-Yakobson-2001}
K.N. Kudin, G.E. Scuseria, and B.I. Yakobson.
\newblock {C2F, BN, and C nanoshell elasticity from ab initio computations}.
\newblock \emph{Phys. Rev. B}, 64\penalty0 (23):\penalty0 235406, 2001.

\bibitem[Huang et~al.(2006)Huang, Wu, and Hwang]{Huang:2006}
Y.~Huang, J.~Wu, and K.C. Hwang.
\newblock {Thickness of graphene and single-wall carbon nanotubes}.
\newblock \emph{Phys. Rev. B}, 74\penalty0 (24):\penalty0 245413, 2006.

\bibitem[Lu(1997)]{Lu:1997}
J.P. Lu.
\newblock {Elastic properties of carbon nanotubes and nanoropes}.
\newblock \emph{Phys. Rev. Lett.}, 79\penalty0 (7):\penalty0 1297--1300, 1997.

\bibitem[Hern\'{a}ndez et~al.(1998)Hern\'{a}ndez, Goze, Bernier, and
  Rubio]{Hernandez:1998}
E.~Hern\'{a}ndez, C.~Goze, P.~Bernier, and A.~Rubio.
\newblock {Elastic properties of C and $B_xC_yN_z$ composite nanotubes}.
\newblock \emph{Phys. Rev. Lett.}, 80\penalty0 (20):\penalty0 4502--4505, 1998.

\bibitem[Arnold(1990)]{Arnold-1990}
D.N. Arnold.
\newblock {Mixed finite element methods for elliptic problems}.
\newblock \emph{Comput. Methods Appl. Mech. Eng.}, 82\penalty0 (1-3):\penalty0
  281--300, 1990.

\bibitem[Miyoshi(1976)]{Miyoshi-1976}
T.~Miyoshi.
\newblock {A mixed finite element method for the solution of the von
  K\'{a}rm\'{a}n equations}.
\newblock \emph{Numer. Math.}, 26\penalty0 (3):\penalty0 255--269, 1976.

\bibitem[Reinhart(1982)]{Reinhart-1982}
L.~Reinhart.
\newblock {On the numerical analysis of the von K\'{a}rm\'{a}n equations: mixed
  finite element approximation and continuation techniques}.
\newblock \emph{Numer. Math.}, 39\penalty0 (3):\penalty0 371--404, 1982.

\bibitem[Blum and Rannacher(1990)]{Blum-Rannacher-1990}
H.~Blum and R.~Rannacher.
\newblock {On mixed finite element methods in plate bending analysis}.
\newblock \emph{Comput. Mech.}, 6\penalty0 (3):\penalty0 221--236, 1990.

\bibitem[Oukit and Pierre(1996)]{Oukit-Pierre-1996}
A.~Oukit and R.~Pierre.
\newblock {Mixed finite element for the linear plate problem: the
  Hermann-Miyoshi model revisited}.
\newblock \emph{Numer. Math.}, 74\penalty0 (4):\penalty0 453--477, 1996.

\bibitem[Pereira et~al.(2009)Pereira, {Castro Neto}, and Peres]{PereiraPRB2009}
V.M. Pereira, A.H. {Castro Neto}, and N.M.R. Peres.
\newblock {Tight-binding approach to uniaxial strain in graphene}.
\newblock \emph{Phys. Rev. B}, 80\penalty0 (4):\penalty0 045401, 2009.

\bibitem[Ribeiro et~al.(2009)Ribeiro, Pereira, Peres, Briddon, and {Castro
  Neto}]{Ribeiro:2009}
R.M. Ribeiro, V.M. Pereira, N.M.R. Peres, P.R. Briddon, and A.H. {Castro Neto}.
\newblock {Strained graphene: tight-binding and density functional
  calculations}.
\newblock \emph{New J. Phys.}, 11\penalty0 (11):\penalty0 115002, 2009.

\bibitem[Ni et~al.(2009)Ni, Yu, Lu, Wang, Feng, and Shen]{Ni:2010}
Z.H. Ni, T.~Yu, Y.H. Lu, Y.Y. Wang, Y.P. Feng, and Z.X. Shen.
\newblock {Uniaxial strain on graphene: Raman spectroscopy study and band-gap
  opening}.
\newblock \emph{ACS Nano}, 3\penalty0 (2):\penalty0 483, 2009.

\bibitem[Farjam and Rafii-Tabar(2009)]{Farjam:2010}
M.~Farjam and H.~Rafii-Tabar.
\newblock {Comment on “Band structure engineering of graphene by strain:
  first-principles calculations”}.
\newblock \emph{Phys. Rev. B}, 80\penalty0 (16):\penalty0 167401, 2009.

\bibitem[Choi et~al.(2010)Choi, Jhi, and Son]{Son:2010}
S.-M. Choi, S.-H. Jhi, and Y.-W. Son.
\newblock {Effects of strain on electronic properties of graphene}.
\newblock \emph{Phys. Rev. B}, 81\penalty0 (8):\penalty0 081407, 2010.

\bibitem[Pereira et~al.(2010)Pereira, Ribeiro, Peres, and {Castro
  Neto}]{Pereira-OpticalStrain:2010}
V.M. Pereira, R.M. Ribeiro, N.M.R. Peres, and A.H. {Castro Neto}.
\newblock {Optical properties of strained graphene}.
\newblock \emph{Europhys. Lett.}, 92\penalty0 (6):\penalty0 67001, 2010.

\bibitem[Hasegawa et~al.(2006)Hasegawa, Konno, Nakano, and
  Kohmoto]{Hasegawa2006Zero}
Y.~Hasegawa, R.~Konno, H.~Nakano, and M.~Kohmoto.
\newblock {Zero modes of tight-binding electrons on the honeycomb lattice}.
\newblock \emph{Phys. Rev. B}, 74\penalty0 (3):\penalty0 033413, 2006.

\bibitem[Ma\~{n}es et~al.(2013)Ma\~{n}es, de~Juan, Sturla, and
  Vozmediano]{Manes2013Generalized}
J.L. Ma\~{n}es, F.~de~Juan, M.~Sturla, and M.A.H. Vozmediano.
\newblock Generalized effective hamiltonian for graphene under nonuniform
  strain.
\newblock \emph{Phys. Rev. B}, 88, 2013.

\bibitem[Ramezani~Masir et~al.(2013)Ramezani~Masir, Moldovan, and
  Peeters]{RamezaniMasir2013Pseudo}
M.~Ramezani~Masir, D.~Moldovan, and F.M. Peeters.
\newblock Pseudo magnetic field in strained graphene: Revisited.
\newblock \emph{Solid State Commun.}, 175:\penalty0 76, 2013.

\bibitem[Zienkiewicz and Zhu(1992)]{Zienkiewicz-Zhu-1992}
O.C. Zienkiewicz and J.Z. Zhu.
\newblock {The superconvergent patch recovery and a posteriori error estimates.
  Part 1: The recovery technique}.
\newblock \emph{Int. J. Numer. Methods Eng.}, 33\penalty0 (7):\penalty0
  1331--1364, 1992.

\bibitem[Tr\"{o}ltzsch(2010)]{Troltzsch-2010}
F.~Tr\"{o}ltzsch.
\newblock \emph{{Optimal Control of Partial Differential Equations: Theory,
  Methods and Applications}}, volume 112 of \emph{Graduate Studies in
  Mathematics}.
\newblock American Mathematical Society, 2010.

\bibitem[Borz\`{\i} and Schulz(2011)]{Borzi-Schulz-2011}
A.~Borz\`{\i} and V.~Schulz.
\newblock \emph{{Computational Optimization of Systems Governed by Partial
  Differential Equations}}.
\newblock Society for Industrial and Applied Mathematics, 2011.

\bibitem[Jones and Mahadevan(2014)]{Jones:OptimalShaping}
G.W. Jones and L.~Mahadevan.
\newblock {Optimal shaping of plates using incompatible strains}.
\newblock \emph{In preparation}, 2014.

\bibitem[Muller et~al.(1980)Muller, Yushchenko, and
  Derjaguin]{Muller-Yushchenko-Derjaguin-1980}
V.M. Muller, V.S. Yushchenko, and B.V. Derjaguin.
\newblock {On the influence of molecular forces on the deformation of an
  elastic sphere and its sticking to a rigid plane}.
\newblock \emph{J. Colloid Interface Sci.}, 77\penalty0 (1):\penalty0 91--101,
  1980.

\bibitem[Xu and Buehler(2010)]{Xu-Buehler-2010}
Z.~Xu and M.J. Buehler.
\newblock {Interface structure and mechanics between graphene and metal
  substrates: a first-principles study}.
\newblock \emph{J. Phys. Condens. Matter}, 22\penalty0 (48):\penalty0 485301,
  2010.

\bibitem[Lu et~al.(2013)Lu, Bao, Su, and Loh]{Lu-Bao-etal-2013}
J.~Lu, Y.~Bao, C.L. Su, and K.P. Loh.
\newblock {Properties of strained structures and topological defects in
  graphene}.
\newblock \emph{ACS Nano}, 7\penalty0 (10):\penalty0 8350--8357, 2013.

\bibitem[Zhang et~al.(2014{\natexlab{b}})Zhang, Huang, Velasco, Myhro,
  Maldonado, Tran, Zhao, Wang, Lee, Liu, Bao, and Lau]{Zhang-Huang-etal-2014}
H.~Zhang, J.-W. Huang, J.~Velasco, K.~Myhro, M.~Maldonado, D.D. Tran, Z.~Zhao,
  F.~Wang, Y.~Lee, G.~Liu, W.~Bao, and C.N. Lau.
\newblock {Transport in suspended monolayer and bilayer graphene under strain:
  a new platform for material studies}.
\newblock \emph{Carbon}, 69:\penalty0 336--341, 2014{\natexlab{b}}.

\bibitem[Bunch et~al.(2007)Bunch, van~der Zande, Verbridge, Frank, Tanenbaum,
  Parpia, Craighead, and McEuen]{Bunch-vanderZande-etal-2007}
J.S. Bunch, A.M. van~der Zande, S.S. Verbridge, I.W. Frank, D.M. Tanenbaum,
  J.M. Parpia, H.G. Craighead, and P.L. McEuen.
\newblock {Electromechanical resonators from graphene sheets}.
\newblock \emph{Science}, 315\penalty0 (5811):\penalty0 490--493, 2007.

\bibitem[Garcia-Sanchez et~al.(2008)Garcia-Sanchez, van~der Zande, {San Paulo},
  Lassagne, McEuen, and Bachtold]{GarciaSanchez-vanderZande-etal-2008}
D.~Garcia-Sanchez, A.M. van~der Zande, A.~{San Paulo}, B.~Lassagne, P.L.
  McEuen, and A.~Bachtold.
\newblock {Imaging mechanical vibrations in suspended graphene sheets}.
\newblock \emph{Nano Lett.}, 8\penalty0 (5):\penalty0 1399--1403, 2008.

\bibitem[Koiter(1966)]{Koiter-1966}
W.T. Koiter.
\newblock {On the nonlinear theory of thin elastic shells}.
\newblock \emph{Proc. K. Ned. Akad. van Wet. Ser. B}, 69:\penalty0 1--54, 1966.

\bibitem[Niordson(1985)]{Niordson-1985}
F.I. Niordson.
\newblock \emph{{Shell Theory}}, volume~29 of \emph{North-Holland Series in
  Applied Mathematics and Mechanics}.
\newblock North-Holland, Amsterdam, 1985.

\bibitem[Helfrich(1973)]{Helfrich-1973}
W.~Helfrich.
\newblock {Elastic properties of lipid bilayers: theory and possible
  experiments}.
\newblock \emph{Zeitschrift f\"{u}r Naturforsch. C}, 28\penalty0 (11):\penalty0
  693--703, 1973.

\bibitem[Wei et~al.(2013)Wei, Wang, Wu, Yang, and Dunn]{Wei-Wang-etal-2013}
Y.~Wei, B.~Wang, J.~Wu, R.~Yang, and M.L. Dunn.
\newblock {Bending rigidity and Gaussian bending stiffness of single-layered
  graphene}.
\newblock \emph{Nano Lett.}, 13\penalty0 (1):\penalty0 26--30, 2013.

\bibitem[Koskinen and Kit(2010)]{Koskinen-Kit-2010}
P.~Koskinen and O.O. Kit.
\newblock {Approximate modeling of spherical membranes}.
\newblock \emph{Phys. Rev. B}, 82\penalty0 (23):\penalty0 235420, 2010.

\bibitem[Ciarlet(1980)]{Ciarlet-1980}
P.G. Ciarlet.
\newblock {A justification of the von K\'{a}rm\'{a}n equations}.
\newblock \emph{Arch. Ration. Mech. Anal.}, 73\penalty0 (4):\penalty0 349--389,
  1980.

\bibitem[Ciarlet(1997)]{Ciarlet-1997}
P.G. Ciarlet.
\newblock \emph{{Mathematical Elasticity Volume II: Theory of Plates}},
  volume~27 of \emph{Studies in Mathematics and Its Applications}.
\newblock Elsevier, Amsterdam, 1997.

\bibitem[Friesecke et~al.(2006)Friesecke, James, and
  M\"{u}ller]{Friesecke-James-Muller-2006}
G.~Friesecke, R.D. James, and S.~M\"{u}ller.
\newblock {A hierarchy of plate models derived from nonlinear elasticity by
  Gamma-convergence}.
\newblock \emph{Arch. Ration. Mech. Anal.}, 180\penalty0 (2):\penalty0
  183--236, 2006.

\end{thebibliography}
\bibliographystyle{unsrtnat}

\end{document}